\RequirePackage[T1]{fontenc}
\documentclass[12pt]{article}

\usepackage{times}

\pdfoutput=1
\usepackage[height=8.85in,width=6.45in]{geometry}

\usepackage{amsmath}
\numberwithin{equation}{section}

\usepackage[utf8]{inputenc}
\usepackage[svgnames]{xcolor}
\usepackage{graphicx}
\usepackage{appendix}

\usepackage{ascmac}
\usepackage{amsmath}
\usepackage{amssymb}
\usepackage{mathtools}
\usepackage{bm}

\usepackage{braket}
\usepackage{cancel}
\usepackage{fancybox}
\usepackage{slashed}

\usepackage{multirow}
\usepackage{tikz}
\usepackage{tikz-cd}
\usetikzlibrary{fit}
\usetikzlibrary{shapes.misc}
\usetikzlibrary{decorations.pathmorphing}

\usepackage{array}
\usepackage{extarrows}
\usepackage{colortbl}
\usepackage{cite}
\usepackage
[
	bookmarks=false,
	colorlinks=true,
	citecolor=refcolor,
	linkcolor=refcolor
]{hyperref}
\usepackage{subfig}
\usetikzlibrary{shapes.geometric}
\usetikzlibrary{positioning}
\usepackage[sectionbib]{chapterbib}
\usepackage[bottom]{footmisc}

\definecolor{Blue}{rgb}{0,0,1}
\definecolor{blue}{rgb}{0.0, 0.4, 1}
\definecolor{darkgreen}{rgb}{0.,0.6,0.}
\definecolor{lightyellow}{rgb}{1.0, 0.95, 0.7}
\definecolor{lightblue}{rgb}{0.7, 0.9, 1.0}
\definecolor{lightpink}{rgb}{1.0, 0.85, 0.95}
\definecolor{lightgreen}{rgb}{0.7, 1.0, 0.4}
\definecolor{refcolor}{rgb}{0.3,0.3,1}

\newcommand*{\gray}[1]{\textcolor{lightgray}{#1}}

\definecolor{colorA}{rgb}{1,0,0}
\definecolor{colorB}{rgb}{1,0.5,0}
\definecolor{colorC}{rgb}{0.8,0.6,0.}
\definecolor{colorD}{rgb}{0.6,0.8,0}
\definecolor{colorE}{rgb}{0,0.65,0}
\definecolor{colorF}{rgb}{0,0.72,0.92}
\definecolor{colorG}{rgb}{0,0,1}
\definecolor{colorH}{rgb}{0.72, 0, 0.92}
\definecolor{colorI}{rgb}{0.92, 0.2, 0.8}

\newcommand*{\colorA}[1]{\textcolor{colorA}{#1}}
\newcommand*{\colorB}[1]{\textcolor{colorB}{#1}}
\newcommand*{\colorC}[1]{\textcolor{colorC}{#1}}
\newcommand*{\colorD}[1]{\textcolor{colorD}{#1}}
\newcommand*{\colorE}[1]{\textcolor{colorE}{#1}}
\newcommand*{\colorF}[1]{\textcolor{colorF}{#1}}
\newcommand*{\colorG}[1]{\textcolor{colorG}{#1}}
\newcommand*{\colorH}[1]{\textcolor{colorH}{#1}}
\newcommand*{\colorI}[1]{\textcolor{colorI}{#1}}

\newcommand*{\bZ}{\mathbb{Z}}
\newcommand*{\bR}{\mathbb{R}}

\newcommand*{\cA}{\mathcal{A}}

\newcommand*{\cD}{\mathcal{D}}
\newcommand*{\cE}{\mathcal{E}}

\newcommand*{\cL}{\mathcal{L}}

\newcommand*{\cN}{\mathcal{N}}

\newcommand*{\KB}{\mathrm{KB}}
\renewcommand*{\ker}{\mathrm{Ker}\,}
\renewcommand*{\mod}{\ \mathrm{mod}\ }

\def\beq#1\eeq{\begin{align}#1\end{align}}

\def\index{\mathop{\mathrm{index}}}
\def\Hom{\mathop{\mathrm{Hom}}}

\DeclareMathOperator{\tr}{tr}
\def\U{\mathrm{U}}
\def\SU{\mathrm{SU}}

\def\SO{\mathrm{SO}}
\def\Sp{\mathrm{Sp}}

\def\Spin{\mathrm{Spin}}

\usepackage{spectralsequences}
\DeclareSseqGroup\tower {} {
	\class(0,0)\foreach \i in {1,...,6} {
		\class(0,\i)
		\structline(0,\i-1,-1)(0,\i,-1)
	}
}
\usepackage{arydshln}

\begin{document}

\begin{titlepage}

\begin{flushright}
IPMU-22-0011\\
TU-1148
\end{flushright}

\vskip 2cm

\begin{center}

{\Large \bfseries
Global anomalies in $\bm{8d}$ supergravity}

\vskip 1cm
Yasunori Lee$^{1,2}$ and Kazuya Yonekura$^3$
\vskip 1cm

\begin{tabular}{ll}
$^1$ & Kavli Institute for the Physics and Mathematics of the Universe (WPI), \\
& University of Tokyo,  Kashiwa, Chiba 277-8583, Japan\vspace{1mm}\\
$^2$ & Department of Physics, Faculty of Science, \\
& University of Tokyo, Bunkyo, Tokyo 113-0033, Japan\vspace{1mm}\\
$^3$ & Department of Physics, Faculty of Science, \\
& Tohoku University, Sendai, Miyagi 980-8578, Japan\\
\end{tabular}

\vskip 1cm

\end{center}

\noindent
We study gauge and gravitational anomalies of fermions and 2-form fields on eight-dimensional spin manifolds.
Possible global gauge anomalies are classified by spin bordism groups $\Omega^{\text{spin}}_9(BG)$ 
which we determine by spectral sequence techniques, and we also identify their explicit generator manifolds. 
It turns out that a fermion in the adjoint representation of any simple Lie group,
and a gravitino in  $8d$ $\cN=1$ supergravity theory, have anomalies. We discuss how a 2-form field, which also appears in supergravity,
produces anomalies which cancel against these fermion anomalies in a certain class of supergravity theories.
In another class of theories, the anomaly of the gravitino is not cancelled by the 2-form field, but by topological
degrees of freedom.  
It gives a restriction on the topology of spacetime manifolds which is not visible at the level of differential-form analysis.

\end{titlepage}

\setcounter{tocdepth}{2}
\tableofcontents

\newpage

\section{Introduction and Summary}

Gauge theories are by definition invariant under gauge transformations,
otherwise they are anomalous and inconsistent.
A simple manifestation of the anomaly is non-invariance of the partition function $Z[A]$
when the gauge field $A$ is transformed to $A^g$,
\begin{equation}\label{eq:introgauge}
	Z[A^g] \neq Z[A].
\end{equation}
Typical sources of anomalies are massless chiral fermions.
As is well known, perturbative anomalies are related to indices of Dirac operators 
in two-higher dimensions~\cite{Alvarez-Gaume:1983ict,Atiyah:1984tf,Alvarez:1984yi},
and the Atiyah-Singer index theorem allows us to describe them in terms of anomaly polynomials.
However, even when such perturbative anomalies are absent,
there can also be non-perturbative (global) gauge transformations which cannot be smoothly deformed to the trivial one,
under which theories are not invariant \cite{Witten:1982fp,Witten:1985xe}.

More generally, there are also fermion anomalies which are not as simply represented as~\eqref{eq:introgauge}.
Current understanding of anomalies is that 
they arise in the definition of the partition function $Z[A]$ itself, rather than its gauge transformations.
In particular, chiral fermions in $d$-dimensions can be realized
as boundary modes of massive bulk theories in $(d+1)$-dimensions,
and the anomaly of the original $d$-dimensional boundary theory is given by the partition function of the $(d+1)$-dimensional bulk theory~\cite{Witten:1999eg,Witten:2015aba,Witten:2019bou}, 
which is called the invertible field theory \cite{Freed:2004yc}. 

Furthermore, this sort of argument is not restricted to the cases of fermionic fields,
but also incorporates the cases of bosonic fields.
For example, we know that
the contribution of a 2-form field to the anomaly of ten-dimensional superstring theories is
very important for Green-Schwarz mechanism of anomaly cancellation \cite{Green:1984sg}.
They are studied at the perturbative level in the past,  but they can also produce non-perturbative global anomalies. 

Unfortunately, due to their conceptual subtleties as well as technical difficulties,
the global anomalies have not been thoroughly investigated for decades,
especially in higher dimensions.
But thanks to the recent developments, they are now within range of analyses,
and this paper aims to obtain new results on the case of eight-dimensional theories.\footnote{
	See e.g. \cite{
		Garcia-Etxebarria:2017crf,
		Monnier:2018nfs,
		Freed:2019sco,
		Hsieh:2020jpj,
		Apruzzi:2020zot,
		Cvetic:2020kuw,
		Montero:2020icj,
		Lee:2020ewl,
		Davighi:2020kok,
		Tachikawa:2021mvw,
		Debray:2021vob,
		Tachikawa:2021mby
	} for a sampling of recent studies of global anomalies in higher dimensions.
}

\bigskip

For the purpose of systematic studies of anomalies, an important point is as follows. 
As we mentioned above in the context of fermions,
the anomalies of a quantum field theory (QFT) are believed to be captured by a one-higher dimensional invertible QFT.
These invertible field~theories are known to be described in terms of bordism groups 
\cite{Kapustin:2014tfa,Kapustin:2014dxa,Freed:2016rqq,Yonekura:2018ufj,Yamashita:2021cao},
and in particular, the information of the global anomalies of fermions in $d$-dimensional $G$ gauge theories on spin manifolds
are encoded in the bordism group $\Omega_{d+1}^{\text{spin}}(BG)$,
where $BG$ is the classifying space of the gauge group $G$.\footnote{
	In this paper we only consider spin structure as the spacetime structure,
	but it would also be interesting to consider other structures such as pin$^{-}$ structure. See \cite{Montero:2020icj} for new constraints in 
	eight and nine dimensions when we take those additional structures into account.
}

Our focus will be on eight-dimensional ($8d$) $\cN=1$ supergravity theories,\footnote{
It would also be interesting to study how the anomalies of 7-branes in non-compact ten-dimensional spaces are cancelled,
possibly by a coupling to bulk RR-fields as discussed in \cite{Freed:2000tt}.
}
where some of the gauge groups $G$ are known to be actually realized despite the possible anomalies.
For instance, many such theories can be realized by F-theory compactification on elliptic K3 
surfaces~\cite{Vafa:1996xn,Morrison:1996na,Morrison:1996pp}, including those with ``frozen'' singularities \cite{Witten:1997bs,Tachikawa:2015wka,Bhardwaj:2018jgp}.\footnote{
	See also \cite{Montero:2020icj,Cvetic:2020kuw,Hamada:2021bbz} and references therein for investigations of possible gauge groups in $8d$ supergravity by bottom-up approaches. 
}
One curious observation on these theories with known string-theory realizations is that, the rank of the total gauge algebra of vector multiplets 
is either $18$, $10$, or $2$. 
We will see in this paper that the structure of anomalies are quite different between these three cases,
due to the difference in the structure of the 2-form field.

Let us recall some facts about the 2-form field.
The field strength 3-form $H$ of
the 2-form field $B$ in $8d$ $\cN=1$ supergravity satisfies the equation of the form
\beq
dH = k_{\rm grav} \left[\cN_{\rm grav} \tr R \wedge R \right] + \sum_{i} k_{G_i}  \left[\cN_{G_i} \tr F_{G_i} \wedge F_{G_i}   \right] +\cdots \label{eq:Heq}
\eeq
where $R$ is the Riemann curvature tensor, $F_G$ is the field strength of the $G$ gauge field, $\cN_{\rm grav}$ and $\cN_G$ are 
appropriate normalization factors such that $\cN_{\rm grav} \tr R \wedge R $ and $\cN_G \tr (F_G \wedge F_G)$
correspond to the characteristic class by the Chern-Weil construction which represent integral cohomology classes, and $k_{\rm grav}$ and $k_G$
are the gravitational and gauge Chern-Simons levels. 

We will show that the gaugino of any simple Lie algebra and the gravitino have anomalies, and hence they must be cancelled by some mechanism. 
The situation in the three cases mentioned above are as follows.
\begin{itemize}
\item rank $18$\,:
$k_{\rm grav}(=1)$ and $k_G$ are both odd for all known examples realized by string theory,
and therefore all the fermion anomalies can be cancelled by the 2-form field. 
\item rank $2$\,:
$k_{\rm grav}=0$ for all known examples,\footnote{
	The possibility of realizing $k_{\rm grav}=1$ is not ruled out at the time of writing,
	and if realized, the details will depend on the parity of $k_G$.
	For the theoretical constraints on the value of $k_{\rm grav}$, see \cite{Kim:2019ths}.
}
and the anomaly of the gravitino cannot be cancelled by the 2-form field. We claim that
a topological 3-form $\bZ_2$ gauge field is responsible for the anomaly cancellation, and discuss its origin when the theory is obtained by 
the compactification of M-theory on Klein bottle. 
\item rank $10$\,:
we still do not have a complete understanding.
This case includes the $\Sp(n)$ gauge algebras which have an additional anomaly compared to other Lie algebras~\cite{Garcia-Etxebarria:2017crf}, 
and also simply-laced Lie algebras with even $k_G$.
\end{itemize}
The rest of the paper is organized as follows.

In Sec.\,\ref{sec:explicit}, we first take a bottom-up approach and
compute the $\eta$-invariants of Dirac operators $\cD_9$ on nine-manifolds $S^1 \times M_8$
for some special eight-manifolds $M_8$ or gauge bundles over them.
The Atiyah-Patodi-Singer index theorem \cite{APS-1} tells us that they are in fact bordism invariants,
and as a result we obtain a partial list of generators of bordism groups $\Omega_9^{\text{spin}}(BG)$
of classifying spaces $BG$ of
connected, simply-connected, compact simple Lie groups $G$.
In particular, we find that there is a universal global gauge anomaly to which fermions in adjoint representation contribute for those gauge groups of interest,
which has not been identified by conventional analyses using homotopy groups $\pi_8(G)$.

In Sec.\,\ref{sec:bordism}, we turn to a top-down approach and
compute various bordism groups $\Omega_9^{\text{spin}}(BG)$
by Atiyah-Hirzebruch spectral sequences and Adams spectral sequences.
The results show that the list obtained in the last section is actually complete,
and correspondingly the possible global gauge anomalies are exhausted by
those of fermions charged under representations considered there.
In addition, we also mention the (non simply-connected) $G=\mathrm{SO}(n)$ case along the way.

In Sec.\,\ref{sec:GSmech}, we examine the anomaly of 2-form fields
and discuss the anomaly cancellation utilizing them,
which can be thought of as an $8d$ analog of the Green-Schwarz mechanism \cite{Green:1984sg}.
This in fact renders some of the apparently-noxious theories of fermions anomaly-free,
including those realized as low-energy effective theories of string theories,
just as in the original version in ten dimensions.

In Sec.\,\ref{sec:TQFT}, we also take a look at some of the exceptions to the above,
namely theories with anomalies which cannot be cancelled by 2-form fields.
We take up one of them and argue that the anomaly is actually canceled in the end, but it requires topological degrees of freedom.

\section{Some manifolds and fermion anomalies}\label{sec:explicit}

In this section, we discuss some concrete examples of global anomalies of Weyl fermions in eight-dimensional $G$ gauge theories.
The analysis in Sec.\,\ref{sec:bordism} will show that these examples in fact exhaust all possible anomalies
for connected, simply-connected, compact simple Lie groups $G$.
 
First of all, fermion anomalies in $d$-dimensions are described by the Atiyah-Patodi-Singer (APS) $\eta$-invariant in $(d+1)$-dimensions~\cite{Witten:1985xe,Witten:2015aba,Witten:2019bou}. 
Let $\cD_{d+1}$ be the relevant Dirac operator in $(d+1)$-dimensions\footnote{
	See \cite{Witten:2019bou} for the details of the construction of $\cD_{d+1}$ from the $d$-dimensional data.
} 
and $\lambda_i$'s be its eigenvalues.
The $\eta$-invariant is
defined as\footnote{
	Strictly speaking, this is not the ``genuine'' $\eta$ but rather what is called $\xi(s=0)$ in the original paper \cite{APS-2},
	but here we follow the conventional nomenclature.
}
\beq
\eta = \frac12 \left( \sum_{\lambda_i\neq0}\frac{\lambda_i}{|\lambda_i|}  + \dim \ker \cD_{d+1} \right).
\label{eq:def-eta}
\eeq
Since the anomalies take the form of $e^{-2\pi i\eta}$, we are interested in the values of $\eta$ modulo $\bZ$.

Now let us focus on the $d=8$ case.
All the examples of 9-manifolds we discuss are of the form $S^1 \times M_8$, where $M_8$ is a closed 8-manifold possibly equipped with a $G$-bundle, 
and $S^1$ is assumed to have the periodic (i.e.  non-bounding) spin structure which gives the nontrivial element of the bordism group $\Omega^\text{spin}_1(\rm{pt})$.
On $S^1 \times M_8$, the Dirac operator is of the form
\beq
\cD_9 = i \gamma^{\tau}\partial_\tau + \cD_8
\eeq
where $\cD_8$ is the Dirac operator on $M_8$, and $\tau$ is the coordinate of $S^1$.
Suppose that $\Psi(\tau,y)$ is an eigenfunction of $\cD_9$, where $y$ is a coordinate system on $M_8$.
Then $\gamma^\tau \Psi(-\tau, y)$ is also an eigenfunction with the opposite-sign eigenvalue.
Thus all nonzero modes $\lambda_i \neq 0$ appear 
in pairs with eigenvalues $\pm |\lambda_i|$, and therefore cancel out in the definition \eqref{eq:def-eta} of $\eta$.
Also, $  \ker \cD_9$ is the space of zero modes of $\cD_9$, and these zero modes need to satisfy $\partial_\tau \Psi(\tau,y)=0$ and $\cD_8 \Psi(\tau, y)=0$
since $(\cD_9)^2= (i \partial_\tau)^2 + (\cD_8)^2$ for non-negative operators $ (i \partial_\tau)^2$ and $ (\cD_8)^2$.
Thus $  \dim \ker \cD_9 =  \dim \ker \cD_8 = \index \cD_8$ modulo $2\bZ$.
As a result,
\beq
\eta =  \frac12 \index \cD_8 \ \mod \bZ
\eeq
and we only need to compute $\index \cD_8 \mod 2$.\footnote{
An intuitive meaning of the anomalies detected by $\index \cD_d \mod 2$ is as follows~\cite{Witten:1985xe}. 
We consider the path integral of the $d$-dimensional theory on $M_d$. Fermions have zero modes with the index given by $\index \cD_d$.
If the index is odd, the path integral measure is not invariant under the fermion parity $(-1)^F$ which is just the $2\pi$ rotation of spacetime.
}

Let $R$ be the Riemann curvature 2-form and $F$ be the field strength 2-form for the $G$-bundle.
Suppose that the fermion is coupled to the $G$-bundle in a representation $\rho$.
Then, the index theorem states that 
\beq
 \index \cD_8  =  \int_{M_8} \hat{A}(R) \tr_\rho \exp \tilde F = 
 \int_{M_8} \left(\frac{1}{24}\tr_\rho \tilde F^4 - \frac{1}{48} p_1 \tr_\rho \tilde F^2  + \frac{7p_1^2 -4p_2}{5760} \dim \rho \right)
 \label{eq:indextheorem}
\eeq
where $\tilde F:=\frac{i}{2\pi}F$, $\hat{A}(R)$ is the A-roof genus,
and $p_i$'s
are the Pontrjagin classes given in terms of $R$, which have degree $4i$.
We now want to consider the following 8-manifolds $M_8$ possibly equipped with $G$-bundles:
\begin{itemize}
\item Quaternionic projective plane $\mathbb{HP}^2$. Its cohomology ring $H^*(\mathbb{HP}^2;\bZ)$ is known to be generated 
by a single generator $x \in H^4(\mathbb{HP}^2;\bZ) = \bZ$
such that $\int_{\mathbb{HP}^2} x^2 =1$.
The Pontrjagin classes are $p_1=2x$ and $p_2 = 7x^2$ respectively,
and therefore the third term of \eqref{eq:indextheorem} vanishes.

\item Bott manifold $B$. The Pontrjagin classes are $p_1=0$ and $p_2=-1440b$ where $b$ is such that $\int_{B}b=1$, 
and therefore the third term of \eqref{eq:indextheorem} integrates to $\dim \rho$.

\item $G$-bundle $P_G \to \mathbb{HP}^2$. 
The base $\mathbb{HP}^2$ has a tautological quaternionic line bundle whose structure group is $\Sp(1) = \SU(2)$,
and $P_G$ is obtained by using a map $\SU(2) \to G$ associated with a simple long root.

\item $G$-bundle $Q_G \to S^4 \times S^4$.
Taking an appropriate map $\SU(2) \times \SU(2) \to G$ discussed below,
we take a bundle over the first (resp. second) $S^4$ with the unit second Chern class for the first (resp. second) $\SU(2)$.
\end{itemize}
For more details on the facts about manifolds $\mathbb{HP}^2$ and $B$ mentioned above, see e.g.~\cite[Sec.\,5]{Freed:2019sco}. 
Let us use these manifolds to deduce some possible anomalies. 

First, recall that the anomaly of 
a gravitino can be described by taking the tensor product of the spinor bundle and $TM_8 - \underline{\bR}$, where
$TM_8$ is the tangent bundle of $M_8$ and $\underline{\bR}$ is the trivial bundle, in place of a $G$-bundle \cite{AlvarezGaume:1983ig}.
Taking $F=R$
correspondingly, one yields $ \tr \tilde F^4 = 2(p_1^2 - 2p_2)$ and
$ \tr \tilde F^2 = 2p_1$, and hence\footnote{
	Another view on the subtraction $-1$ is that, it takes the contribution from ghosts into account,
	which amounts to removing that of a singlet fermion.
}
\beq
 \index \cD_8 =  \int_{M_8} \left( \frac{p_1^2-4p_2}{24} +  \frac{7p_1^2 -4p_2}{5760} (8 -1) \right)
 = \int_{M_8} \frac{289p_1^2 -988p_2}{5760}.
\eeq
From this result,
we see that the index for a gravitino is $-1$ on $\mathbb{HP}^2$ and $247$ on $B$, both of which are $1 \mod 2$.

Next, consider the bundle $P_G \to \mathbb{HP}^2$ 
constructed from a quaternionic $\Sp(1)=\SU(2)$ bundle.
In the fundamental representation $\bm{2}$ of $\SU(2)$, we have $ \tr_{\bf{2}} \tilde F^4 = 2x^2$ and $ \tr_{\bf{2}} \tilde F^2 = 2x$
in terms of $x \in H^4(\mathbb{HP}^2 ;\bZ)$,
and thus $ \index \cD_8  =0$.
On the other hand, in the adjoint representation $\bm{3}$,
we have $ \tr_{\bf{3}} \tilde F^4 = 32x^2$ and $ \tr_{\bf{3}} \tilde F^2 = 8x$, and thus $ \index \cD_8  =1$.
Under the map $\SU(2) \to G$ associated with a simple long root,
the adjoint representation adj$(G)$ of generic $G$ decomposes as\footnote{In general, when we have a map (homomorphism) between groups $H \to G$,
a representation of $G$ decomposes into representations of $H$.}
\beq
\textrm{adj}(G) \to \underbrace{\bm{3}}_{\index \cD_8  = 1} \oplus\ 
\underbrace{\bm{2} \oplus \cdots \oplus \bm{2}}_{\index \cD_8  = 0}\ \oplus\ 
\underbrace{\bm{1} \oplus \cdots \oplus \bm{1}}_{\index \cD_8  = 0}
\eeq
and as a result we get $ \index \cD_8  =1$ for adj$(G)$.
This is universal in the sense that it is true
for any compact simple Lie group $G$.

Finally, let us take up the bundle  $Q_G \to S^4 \times S^4$. This was already discussed in \cite{Garcia-Etxebarria:2017crf},
and here we briefly review the argument.
Under the map $ \SU(2) \times \SU(2) \to G$ which we explain in a moment, suppose that a representation $\rho$ of $G$ decomposes as
\beq
\rho \to (\bm{2} \otimes \bm{2})^{\oplus \,\text{odd}}
\oplus (\rho_1 \otimes \bm{1}) \oplus \cdots
\oplus (\bm{1} \otimes \rho_2) \oplus \cdots.
\label{eq:extraanomaly}
\eeq
This condition is satisfied,
for example, in the following cases. 
\begin{itemize}
	\item $\Spin(n \geq 4)$ has a subgroup $\SU(2) \times \SU(2) = \Spin(4) \subset \Spin(n)$.
	The fundamental representation $\bf{n}$ of $\Spin(n)$ satisfies \eqref{eq:extraanomaly} for any $n\geq 4$. 
	The adjoint representation of $\Spin(n)$ also satisfies \eqref{eq:extraanomaly} if $n$ is odd.
	
	\item $\Sp(n \geq 2)$ has a subgroup $\SU(2) \times \SU(2) \subset \Sp(2) \subset \Sp(n)$.
	The adjoint representation satisfies \eqref{eq:extraanomaly}.
	The antisymmetric representation $\bf{n(2n-1)}$ also satisfies \eqref{eq:extraanomaly}.
	
	\item $F_4$ has a subgroup $\Spin(9) \subset F_4$ under which the adjoint representation decomposes as
	$\textrm{adj}(F_4) \to \textrm{adj}(\Spin(9)) \oplus \bf{2^4}$, where $\bf{2^4}$ is the spinor representation of $\Spin(9)$.
	Embedding $\SU(2) \times \SU(2) $ into $\Spin(9)$, the adjoint representation satisfies \eqref{eq:extraanomaly}. 
	The 26-dimensional representation decomposes as $\bf{26}\to\bf{9} \oplus {\bf 2^4} \oplus \bf{1}$ of $\Spin(9)$ and it satisfies \eqref{eq:extraanomaly}.
	
	\item $G_2$ has a subgroup $\tfrac{\SU(2) \times \SU(2)}{\bZ_2} \subset G_2$ and hence we have a map 
	$\SU(2) \times \SU(2)  \to  G_2$.
	The 7-dimensional representation decomposes as 
	$\bf{7} \to (\bf{2} \otimes \bf{2}) \oplus (\bf{1} \otimes \bf{3})$
	and hence satisfies \eqref{eq:extraanomaly}.
\end{itemize}  
For these groups and representations, we get $ \index \cD_8  =1$.
We remark that the adjoint representation of $G_2$ has $ \index \cD_8=0 \mod 2$ for the bundle $Q_G$ studied here.

The results are summarized in Table~\ref{table:pure}.
Notice that all the representations discussed above are real, so there are no perturbative anomalies for fermions charged under them,
and the anomalies detected by $\index \cD_8 \mod 2$ are all global anomalies.
Correspondingly, the $\eta$-invariants become bordism invariants as inferred from the index theorem,
and from Table~\ref{table:pure} we see that
\beq
\renewcommand{\arraystretch}{1.2}
\begin{array}{lcll}
	\bZ_2^{\oplus 2} & \subset & \Omega^\text{spin}_9(\rm{pt})\\ 
	\bZ_2 & \subset & \widetilde \Omega^\text{spin}_9(BG) & \quad (G=\SU(n), E_{6,7,8})\\
	\bZ_2^{\oplus 2} & \subset & \widetilde \Omega^\text{spin}_9(BG) & \quad (G=\Spin(n\geq 4), \Sp(n\geq 2), F_4, G_2)
\end{array}
\eeq
where $\widetilde\Omega^\text{spin}_\bullet$ is the reduced spin bordism group 
(i.e. $\Omega^\text{spin}_\bullet(X) = \widetilde\Omega^\text{spin}_\bullet(X) \oplus \Omega^\text{spin}_\bullet(\rm{pt})$).
It is known that $ \Omega^\text{spin}_9(\rm{pt}) = \bZ_2^{\oplus 2} $,
and by using spectral sequences, we will further show that the manifolds and bundles discussed above exhaust all the generators of $\widetilde\Omega^\text{spin}_9(BG) $
for connected, simply-connected, compact simple Lie groups $G$ in the next section.

\bigskip

\begin{table}[h]
\caption{$\index \cD_8 \mod 2$ on various manifolds for the fermion representations discussed in the main text.
For $\Spin(n)$ we only consider $n\geq 4$, and for $\Sp(n)$ we only consider $n\geq 2$.
}\label{table:pure}
\centering
\renewcommand{\arraystretch}{1.2}
\begin{tabular}{|c|c|c|c|c|}
\hline
\rowcolor{lightyellow}
$ \index \cD_8 \mod 2$ & $\mathbb{HP}^2$ & $B$ & $P_G \to \mathbb{HP}^2$ & $Q_G \to S^4 \times S^4$ \\ \hline
singlet fermion & 0 & 1 & $-$ & $-$ \\ \hline
gravitino & 1 & 1 & $-$ & $-$ \\ \hline 
\multicolumn{1}{|l|}{$\textrm{adj}(G)$~~($ \SU(n), \Spin(2n), E_{6,7,8}, G_2)$} & 0 & $\dim \textrm{adj}(G) \mod 2$ & 1 & 0 if defined \\ \hline
\multicolumn{1}{|l|}{$\textrm{adj}(G)$~~($\Spin(2n+1), \Sp(n), F_4$)} & 0 & $\dim \textrm{adj}(G) \mod 2$ & 1 & 1 \\ \hline
$\begin{tabular}{ccl}
	$\bf{n}$ & of & $\Spin(n)$\\
	$\bf{n(2n-1)}$ & of & $\Sp(n)$\\
	$\bf{26}$ & of & $F_4$\\
	$\bf{7}$ & of & $G_2$
\end{tabular}$
& 0 & $\dim \rho\ \mod 2$ & 0 & 1 \\ \hline
\end{tabular}
\end{table}

\newpage

\section{Bordism group computation}\label{sec:bordism}

In this section, we compute the spin bordism group $\Omega_{d+1}^{\text{spin}}(BG)$
for some simple Lie groups $G$'s.
Roughly speaking, it is a group formed by equivalence classes of closed manifolds equipped with spin structure and $G$-bundle,
where two manifolds are defined to be equivalent if there is a one-higher dimensional compact manifold connecting them.
It can be computed using various types of spectral sequences;
for general introduction to spectral sequences see e.g.~\cite{Hatcher, DavisKirk, McCleary:UsersGuide},
while we also refer to
\cite{Garcia-Etxebarria:2018ajm} for the introduction to Atiyah-Hirzebruch spectral sequences aimed at physicists,
and \cite{BCGuide} for the introduction to Adams spectral sequences.

\subsection{Atiyah-Hirzebruch spectral sequence}

For the Atiyah-Hirzebruch spectral sequence  associated with the trivial fibration
\begin{equation}
	\mathrm{pt}
	\longrightarrow X
	\longrightarrow X,
\end{equation}
the $E^2$-terms are given by
ordinary homology groups $H_p(X; \Omega_q^{\text{spin}}(\mathrm{pt}))$,
and it converges to the bordism group $\Omega_{p+q}^{\text{spin}}(X)$.

\subsubsection{$\mathrm{SU}(n)$ gauge anomaly}
Let us first carry out an explicit computation for the $X=B\mathrm{SU}(n\geq 5)$ case.
The (co)homology of $B\mathrm{SU}(n)$ is known to be
\begin{equation}
	H^\ast(B\mathrm{SU}(n); \bZ)
	=
	\bZ[c_2, c_3,c_4,c_5, \ldots],
\end{equation}
where $c_i \in H^{2i}(B\mathrm{SU}(n); \bZ) $ are Chern classes.
One can easily fill in the $E^2$-page of the Atiyah-Hirzebruch spectral sequence as follows:
\begin{equation}
	\begin{array}{c}
		E^2_{p,q}=H_p\big(B\mathrm{SU}(n);\Omega^{\text{spin}}_q(\mathrm{pt})\big)\vspace{3mm}\\
		\begin{array}{c|c:cccccccccccc}
			9 & \bZ_2^{\oplus 2}\cellcolor{lightyellow} &&&& \ast && \ast && \ast && \ast\\
			8  & \bZ^{\oplus 2} & \cellcolor{lightyellow} &&& \ast && \ast && \ast && \ast\\
			7  &&& \cellcolor{lightyellow}\\
			6  &&&& \cellcolor{lightyellow}\\
			5  &&&&& \cellcolor{lightyellow}\\
			4  & \bZ &&&& \bZ & \cellcolor{lightyellow} & \ast && \ast && \ast\\
			3  & \phantom{\bZ_2} & \phantom{\bZ_2} & \phantom{\bZ_2} & \phantom{\bZ_2} & \phantom{\bZ_2} & \phantom{\bZ_2} & \phantom{\bZ_2} \cellcolor{lightyellow} & \phantom{\bZ_2} & \phantom{\bZ_2} & \phantom{\bZ_2} & \phantom{\bZ_2}\\
			2  & \bZ_2 &&&& \bZ_2 && \bZ_2 & \cellcolor{lightyellow} & \ast && \ast\\
			1  & \bZ_2 &&&& \bZ_2 && \bZ_2 && \cellcolor{lightyellow}\bZ_2^{\oplus 2} && \ast\\
			0 & \bZ &&&& \bZ && \bZ && \bZ^{\oplus 2} & \cellcolor{lightyellow} & \bZ^{\oplus 2} \\
			\hline
			& 0 & 1 & 2 & 3 & 4 & 5 & 6 & 7 & 8 & 9 & 10\\
		\end{array}
	\end{array}
\end{equation}
where the horizontal and vertical axes correspond to $p$ and $q$ respectively.

Here, the differentials $d^2: E^2_{p,q} \to E^2_{p-2,q+1}$ for $q=0,1$
are known \cite{Teichner1993} to be the duals of the Steenrod square $Sq^2$
(composed with mod-2 reduction for $q=0$).
From the knowledge on the cohomology of $B\mathrm{SU}(n)$,
one can confirm that $d^2: E^2_{10,0} \to E^2_{8,1}$ is non-trivial since $Sq^2 c_4 = c_5$
for mod-2 reduced Chern classes,
while $d^2: E^2_{8,1} \to E^2_{6,2}$ is trivial,
and also $d^4: E^2_{8,1} \to E^2_{4,4}$ is obviously trivial as it should be a homomorphism.
As a result, one is led to
\begin{equation}
	\widetilde \Omega_9^{\text{spin}}(B\mathrm{SU}(n)) = \bZ_2
\end{equation}
and this detects the universal anomaly of adjoint fermions in $8d$ $\mathrm{SU}(n)$ gauge theories described in the previous section.

\subsubsection{$\mathrm{Sp}(n)$ gauge anomaly}
Similarly, for $X=B\mathrm{Sp}(n\geq 2)$, it is known that the (co)homology is
\begin{equation}
	H^\ast(B\mathrm{Sp}(n); \bZ)
	=
	\bZ[q_1, q_2, \ldots],
\end{equation}
where $q_{i} \in H^{4i}(B\mathrm{Sp}(n); \bZ)$.
One can again easily fill in the $E^2$-page of the Atiyah-Hirzebruch spectral sequence as follows:
\begin{equation}
	\begin{array}{c}
		E^2_{p,q}=H_p\big(B\mathrm{Sp}(n);\Omega^{\text{spin}}_q(\mathrm{pt})\big)\vspace{3mm}\\
		\begin{array}{c|c:cccccccccccc}
			9 & \bZ_2^{\oplus 2}\cellcolor{lightyellow} &&&& \ast &&&& \ast && \\
			8  & \bZ^{\oplus 2} & \cellcolor{lightyellow} &&& \ast &&&& \ast &&\\
			7  &&& \cellcolor{lightyellow}\\
			6  &&&& \cellcolor{lightyellow}\\
			5  &&&&& \cellcolor{lightyellow}\\
			4  & \bZ &&&& \bZ & \cellcolor{lightyellow} &&& \ast &&\\
			3  & \phantom{\bZ_2} & \phantom{\bZ_2} & \phantom{\bZ_2} & \phantom{\bZ_2} & \phantom{\bZ_2} & \phantom{\bZ_2} & \phantom{\bZ_2} \cellcolor{lightyellow} & \phantom{\bZ_2} & \phantom{\bZ_2} & \phantom{\bZ_2} & \phantom{\bZ_2}\\
			2  & \bZ_2 &&&& \bZ_2 &&& \cellcolor{lightyellow} & \ast && \\
			1  & \bZ_2 &&&& \bZ_2 &&&& \cellcolor{lightyellow}\bZ_2^{\oplus 2} && \\
			0 & \bZ &&&& \bZ &&&& \bZ^{\oplus 2} & \cellcolor{lightyellow} & \\
			\hline
			& 0 & 1 & 2 & 3 & 4 & 5 & 6 & 7 & 8 & 9 & 10\\
		\end{array}
	\end{array}
\end{equation}
Since $E_{10,0}^2$ is empty opposed to the $X=B\mathrm{SU}(n \geq 5)$ case,
one is led to conclude that
\begin{equation}
	\widetilde \Omega_9^{\text{spin}}(B\mathrm{Sp}(n)) = \bZ_2^{\oplus 2}
\end{equation}
where the additional $\bZ_2$ should correspond to the subtler anomaly
discussed in \cite{Garcia-Etxebarria:2017crf}.

However, it is not always the case that the Atiyah-Hirzebruch spectral sequence is adequate
to obtain the desired bordism groups.
In the next subsection, we will introduce another spectral sequence
which can be further exploited in such cases.

\subsection{Adams spectral sequence}

For the case of interest, the $E_2$-terms of the Adams spectral sequence are given as
\begin{equation}
	E_2^{s,t}
	=
	\mathrm{Ext}_{\cA}^{s,t}\big(\widetilde H^\ast(M\mathrm{Spin}\wedge X;\bZ_2), \bZ_2)
	\ \Longrightarrow\ 
	\pi^\mathrm{st}_{t-s}(M\mathrm{Spin} \wedge X)^{\wedge}_2
	\ \simeq\ 
	\widetilde\Omega_{t-s}^{\text{spin}}(X)^{\wedge}_2,
\end{equation}
and converge to the 2-completion of a stable homotopy group,
which is isomorphic to that of the desired (reduced) bordism group via the Pontrjagin-Thom construction.
Here, $\cA$ is the mod-2 Steenrod algebra generated by certain cohomology operations,
$\mathrm{Ext}_{R}$ is a certain functor in the category of (graded) $R$-modules
which takes values in Abelian groups,
and $M\mathrm{Spin}$ is the Thom spectrum of the universal bundle over $B\mathrm{Spin}$.

Using the K\"unneth formula, the (reduced) cohomology of a smash product is decomposed as
\begin{equation}
	\widetilde H^\ast(X\wedge Y;\bZ_2)
	\simeq
	\widetilde H^\ast(X;\bZ_2)
	\otimes_{\bZ_2}
	\widetilde H^\ast(Y;\bZ_2).
\end{equation}
Note that it is known \cite{ABP1967, Freed:2016rqq, MengGuoThesis} that
\begin{equation}
	\widetilde H^\ast(M\mathrm{Spin};\bZ_2)
	\simeq
	\cA \otimes_{\cA(1)} (\bZ_2 \oplus \Sigma^8\bZ_2 \oplus \Sigma^{10} J \oplus M_{\geq 16})
\end{equation}
where $\cA(1)$ is the subalgebra of $\cA$ generated by $Sq^1$ and $Sq^2$,
$J$ is a certain $\cA(1)$-module called the \textit{joker},
and $M_{\geq 16}$ is also an $\cA(1)$-module which is trivial in degrees less than $16$.
Then, the combination of the shearing isomorphism and the adjunction formula
allows us to rewrite the $E_2$-terms as
\begin{equation}
	\mathrm{Ext}_{\cA(1)}^{s,t}\big(
		(\bZ_2 \oplus \Sigma^8\bZ_2 \oplus \Sigma^{10} J \oplus M_{\geq 16})
		\otimes_{\bZ_2}
		\widetilde H^\ast(X;\bZ_2), \bZ_2).
\end{equation}
Fortunately, for simply-connected compact simple Lie groups $G$,
things become significantly easier
since
the lowest degree of elements in $\widetilde H^\ast(BG;\bZ)$ is $4$,
meaning that for $t-s \leq 11$ the $E_2$-terms can actually be reduced to
\begin{equation}
	\mathrm{Ext}_{\cA(1)}^{s,t}\big(
		\bZ_2
		\otimes_{\bZ_2}
		\widetilde H^\ast(X;\bZ_2), \bZ_2
	\big)
	=
	\mathrm{Ext}_{\cA(1)}^{s,t}\big(
		\widetilde H^\ast(X;\bZ_2), \bZ_2
	\big)
\end{equation}
which converges to the (reduced) $ko$ group.
Therefore, for such $G$ we have
\begin{equation}
	\widetilde\Omega_{d}^{\text{spin}}(BG)
	\simeq
	\widetilde{ko}_{d}(BG)
	\quad
	\text{for }d\leq 11.
\end{equation}

\subsection{$G_2$ gauge anomaly}
Now, let us look at the $X=BG_2$ case.
It is known \cite{Borel1954, MimuraToda} that
the cohomology of $BG_2$ is $p$-torsion free for $p\geq 3$,
which assures that the full bordism group can be derived from its 2-completion.
Furthermore, the $\bZ_2$ cohomology ring is given as
\begin{equation}
	H^\ast(BG_2;\bZ_2) = \bZ_2[y_4, y_6, y_7],
\end{equation}
and the cohomology operations act as
\begin{equation}
	\begin{array}{lcc}
		Sq^2 y_4 & = & y_6,\\
		Sq^1 y_6 & = & y_7.
	\end{array}
\end{equation}
Then, the $\cA(1)$-module structure of $H^\ast(BG_2;\bZ_2)$ for the range of interest is represented as
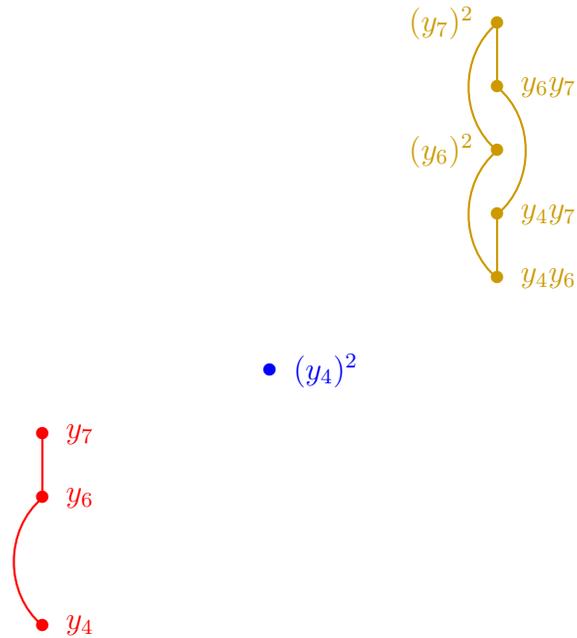
\begin{equation}
	\begin{tikzpicture}[thick]
		\matrix (m) [
			matrix of math nodes,
			row sep= 0.9em,
			column sep=6em
		]{
								&						& \colorC{\bullet}\\
								&						& \colorC{\bullet}\\
								&						& \colorC{\bullet}\\
								&						& \colorC{\bullet}\\
								&						& \colorC{\bullet}\\
								&						& \\
								& \colorG{\bullet} & \\
			\colorA{\bullet}\\
		    \colorA{\bullet}\\
			\phantom{\bullet}\\
			\colorA{\bullet} \\
		};
		\draw[colorA] (m-8-1.center) to (m-9-1.center);
		\draw[colorA] (m-9-1.center) to [out=220, in = 140] (m-11-1.center);
		\draw[colorC] (m-1-3.center) to (m-2-3.center);
		\draw[colorC] (m-4-3.center) to (m-5-3.center);
		\draw[colorC] (m-3-3.center) to [out=220, in = 140] (m-5-3.center);
		\draw[colorC] (m-1-3.center) to [out=220, in = 140] (m-3-3.center);
		\draw[colorC] (m-2-3.center) to [out=320, in = 40] (m-4-3.center);
		\node[colorA, anchor = west, inner sep=3mm] at (m-11-1) {$y_4$};
		\node[colorA, anchor = west, inner sep=3mm] at (m-9-1) {$y_6$};
		\node[colorA, anchor = west, inner sep=3mm] at (m-8-1) {$y_7$};
		\node[colorG, anchor = west, inner sep=3mm] at (m-7-2) {$(y_4)^2$};
		\node[colorC, anchor = west, inner sep=3mm] at (m-5-3) {$y_4y_6$};
		\node[colorC, anchor = west, inner sep=3mm] at (m-4-3) {$y_4y_7$};
		\node[colorC, anchor = east, inner sep=3mm] at (m-3-3) {$(y_{6})^2$};
		\node[colorC, anchor = east, inner sep=3mm] at (m-1-3) {$(y_{7})^2$};
		\node[colorC, anchor = west, inner sep=3mm] at (m-2-3) {$y_6y_7$};
	\end{tikzpicture}
\end{equation}
where the straight lines and curved lines represent the actions of $Sq^1$ and $Sq^2$ respectively. 
Namely, as an $\cA(1)$-module one has
\begin{equation}
	\colorA{\Sigma^4 Q}
	\oplus
	\colorG{\Sigma^8 \bZ_2}
	\oplus
	\colorC{\Sigma^{10} J}
\end{equation}
where $Q$ and $J$ are the ``named'' $\cA(1)$-modules.
Correspondingly, the associated Adams chart
which pictorially describes the $E_2$-page 
$\mathrm{Ext}_{\cA(1)}^{s,t}\big(
	\widetilde H^\ast(BG_2;\bZ_2), \bZ_2
\big)$ is given \cite{BCGuide} by
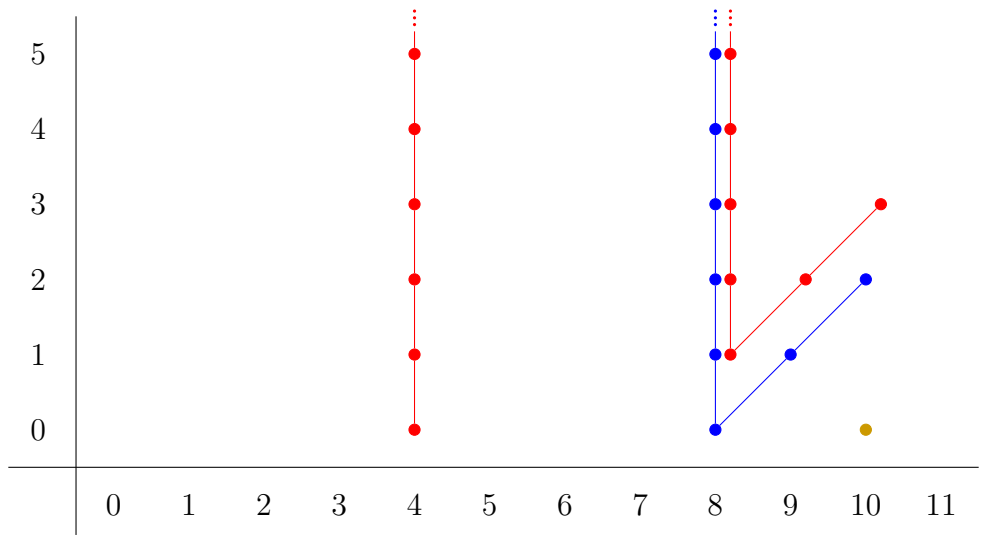
\begin{center}
\begin{sseqdata}[
	name=MBG2,
	Adams grading,
	classes = fill,
	xrange = {0}{11},
	yrange = {0}{5}
]
	\tower[colorA](4,0)
	\tower[opacity=0](8,0)
	\tower[colorG](8,0)
	\class[opacity=0](8,0)    
	\tower[colorA](8,1)
	\class[opacity=0](9,1)
	\class[colorG](9,1)
	\class[opacity=0](9,1)
	\structline[colorG](8,0,2)(9,1,2)
	\class[opacity=0](10,2)
	\class[colorG](10,2)
	\class[opacity=0](10,2)
	\structline[colorG](9,1,2)(10,2,2)
	\class[opacity=0](9,2)
	\class[opacity=0](9,2)
	\class[colorA](9,2)
	\structline[colorA](8,1,3)(9,2,3)
	\class[opacity=0](10,3)
	\class[opacity=0](10,3)
	\class[colorA](10,3)
	\structline[colorA](9,2,3)(10,3,3)
	\class[colorC, opacity=1](10,0)
\end{sseqdata}
\printpage[name = MBG2, page=2]
\end{center}
where the horizontal and vertical axes correspond to $t-s$ and $s$ respectively.
Here, the dots denote the $\bZ_2$-generators in $\mathrm{Ext}_{\cA(1)}^{s,t}(H^\ast(BG_2;\bZ_2), \bZ_2)$,
while the vertical (resp.~sloped) lines represent the action by 
$h_0\in \mathrm{Ext}_{\cA(1)}^{1,1}(\bZ_2, \bZ_2)$
(resp.~$h_1\in \mathrm{Ext}_{\cA(1)}^{1,2}(\bZ_2, \bZ_2)$).
The possibly-nontrivial differentials are the ones with the source at $(t-s,s)=(10,0)$
and would hit the classes in $t-s=9$,
but such differentials are not consistent with the action of $h_1$,
and therefore there are no differentials at all.
As a result, the Adams spectral sequence converges as follows:
\begin{equation}
	\label{omegaBG2}
	\renewcommand{\arraystretch}{1.3}
	\begin{array}{c|ccccccccccccccccc}
		d & 0 & 1 & 2 & 3 & 4 & 5 & 6 & 7 & 8 & 9 & 10 & 11\\
		\hline
		\widetilde\Omega_{d}^{\text{spin}}(BG_2)
		& 0 & 0 & 0 & 0 & \bZ & 0 & 0 & 0 & \bZ^{\oplus 2} & \bZ_2^{\oplus 2} & \bZ_2^{\oplus 3} & 0\\
		\hline
	\end{array}
\end{equation}

\bigskip

\noindent
Comparing the degree-9 part with the Atiyah-Hirzebruch spectral sequence
\begin{equation}
	\begin{array}{c}
		E^2_{p,q}=H_p\big(BG_2;\Omega^{\text{spin}}_q\big)\vspace{3mm}\\
		\begin{array}{c|c:cccccccccccc}
			9 & \bZ_2^{\oplus 2}\cellcolor{lightyellow} &&&& \ast && \ast & \ast & \ast\\
			8  & \bZ^{\oplus 2} & \cellcolor{lightyellow} &&& \ast && \ast && \ast\\
			7  &&& \cellcolor{lightyellow}\\
			6  &&&& \cellcolor{lightyellow}\\
			5  &&&&& \cellcolor{lightyellow}\\
			4  & \bZ &&&& \bZ & \cellcolor{lightyellow} & \ast && \ast\\
			3  &&&&&&& \cellcolor{lightyellow}\\
			2  & \bZ_2 &&&& \gray{\bZ_2} && \bZ_2 & \cellcolor{lightyellow} \bZ_2 & \ast &\\
			1  & \bZ_2 & \phantom{\bZ_2} & \phantom{\bZ_2} & \phantom{\bZ_2} & \gray{\bZ_2} & \phantom{\bZ_2} & \gray{\bZ_2} & \bZ_2 & \cellcolor{lightyellow}\bZ_2 & \phantom{\bZ_2}\\
			0 & \bZ &&&& \bZ && \gray{\bZ_2} && \bZ & \cellcolor{lightyellow}\\
			\hline
			& 0 & 1 & 2 & 3 & 4 & 5 & 6 & 7 & 8 & 9 \\
		\end{array}
	\end{array}
\end{equation}
one indeed notices that there is another type of
global (non-perturbative) gauge anomaly for gauge group $G_2$
which corresponds to the $E^2_{7,2}$,
in addition to the universal one corresponding to the $E^2_{8,1}$.
We claim this to be the traditional anomaly captured by the homotopy group $\pi_8(G_2)$.
The fact that the representation $\bf{7}$ has the anomaly associated to $\pi_8(G_2)$ has been shown in \cite{Garcia-Etxebarria:2017crf}.

\subsection{$F_4$ gauge anomaly}
Similar but a little more complicated case is $X = BF_4$.
The mod-2 (co)homology is known to be
\begin{equation}
	H^\ast(BF_4; \bZ_2)
	=
	\bZ[y_4, y_6, y_7, y_{16}, y_{24}]
\end{equation}
where the action of cohomology operations are the same as $BG_2$
\begin{equation}
	\begin{array}{lcc}
		Sq^2 y_4 & = & y_6,\\
		Sq^1 y_6 & = & y_7,
	\end{array}
\end{equation}
which leads to the same analysis on the Adams spectral sequence for the range of interest.
However, this time there are 3-torsions \cite{TodaBF4} (but $p$-torsion free for $p\geq 5$),
and correspondingly the $E^2$-page of the Atiyah-Hirzebruch spectral sequence becomes
\begin{equation}
	\begin{array}{c}
		E^2_{p,q}=H_p\big(BF_4;\Omega^{\text{spin}}_q\big)\vspace{3mm}\\
		\begin{array}{c|c:cccccccccccc}
			9 & \bZ_2^{\oplus 2}\cellcolor{lightyellow} &&&& \ast && \ast & \ast & \ast\\
			8  & \bZ^{\oplus 2} & \cellcolor{lightyellow} &&& \ast && \ast && \ast\\
			7  &&& \cellcolor{lightyellow}\\
			6  &&&& \cellcolor{lightyellow}\\
			5  &&&&& \cellcolor{lightyellow}\\
			4  & \bZ &&&& \bZ & \cellcolor{lightyellow} & \ast && \ast\\
			3  &&&&&&& \cellcolor{lightyellow}\\
			2  & \bZ_2 &&&& \gray{\bZ_2} && \bZ_2 & \cellcolor{lightyellow} \bZ_2 & \ast &\\
			1  & \bZ_2 & \phantom{\bZ_2} & \phantom{\bZ_2} & \phantom{\bZ_2} & \gray{\bZ_2} & \phantom{\bZ_2} & \gray{\bZ_2} & \bZ_2 & \cellcolor{lightyellow}\bZ_2 & \phantom{\bZ_2}\\
			0 & \bZ &&&& \bZ && \gray{\bZ_2} && \bZ\oplus \bZ_3 & \cellcolor{lightyellow}\\
			\hline
			& 0 & 1 & 2 & 3 & 4 & 5 & 6 & 7 & 8 & 9 \\
		\end{array}
	\end{array}
\end{equation}
Fortunately, the 3-torsion part is irrelevant for our purpose as one can read off,
and in the same way as discussed in the $G_2$ case,
there is an additional $\bZ_2$ in $E_{7,2}^2$ which should correspond to the traditional anomaly captured by $\pi_8(F_4)$.
Again, it is known that the adjoint representation of $F_4$ has the anomaly associated to $\pi_8(F_4)$ \cite{Garcia-Etxebarria:2017crf}.

\subsection{$E_{6,7,8}$ gauge anomaly (and 2-form fields)}

For our purpose, the classifying spaces $BE_{6,7,8}$ can be identified with the Eilenberg-MacLane space $K(\bZ,4)$,
since they are homotopically equivalent at the range of interest \cite{BottSamelson, Kachi1968} as
\begin{equation}
	\label{homotopy-E6E7E8}
	\renewcommand{\arraystretch}{1.3}
	\begin{array}{c|ccccccccccccccccccccccccccccccc}
		d & 1 & 2 & 3 & 4 & 5 & 6 & 7 & 8 & 9 & 10 & 11 & 12 & 13 & 14 & 15 & 16 & \cdots\\
		\hline
		\pi_d(BE_6) & 0 & 0 & 0 & \bZ & 0 & 0 & 0 & 0 & 0 & \bZ\cellcolor{black!10} & 0\cellcolor{black!10} & \bZ\cellcolor{black!10} & \bZ_4\cellcolor{black!10} & 0\cellcolor{black!10} & 0\cellcolor{black!10} & \bZ\cellcolor{black!10} & \cdots\cellcolor{black!10}\\
		\pi_d(BE_7) & 0 & 0 & 0 & \bZ & 0 & 0 & 0 & 0 & 0 & 0 & 0 & \bZ\cellcolor{black!10} & \bZ_2\cellcolor{black!10} & \bZ_2\cellcolor{black!10} & 0\cellcolor{black!10} & \bZ\cellcolor{black!10} & \cdots\cellcolor{black!10}\\
		\pi_d(BE_8) & 0 & 0 & 0 & \bZ & 0 & 0 & 0 & 0 & 0 & 0 & 0 & 0 & 0 & 0 & 0 & \bZ\cellcolor{black!10} & \cdots\cellcolor{black!10}\\
		\hline
	\end{array}
	\setlength{\belowdisplayskip}{20pt}
\end{equation}
Although the situation is complicated since $K(\bZ,4)$ is known to have 3-torsion in the cohomology as $BF_4$ does,
one can nevertheless compute the bordism groups \cite{StongAppendix, JohnFrancis2005}.

The $\bZ_2$-cohomology is known \cite{Serre1953} to be
\begin{equation}
	H^\ast(K(\bZ,4);\bZ_2) = \bZ_2[u_4, u_6, u_7, u_{10}, u_{11}, u_{13}, \ldots]
\end{equation}
where
\begin{equation}
	\begin{array}{lccclccclcc}
		Sq^2 u_4 & = & u_6, && Sq^4 u_6 & = & u_{10}, && Sq^1 u_{10} & = & u_{11},\\
		Sq^1 u_6 & = & u_7, && Sq^6 u_7 & = & u_{13}, && Sq^2 u_{11} & = & u_{13}.\\
	\end{array}
\end{equation}
Then, the relevant $\cA(1)$-module can be represented as
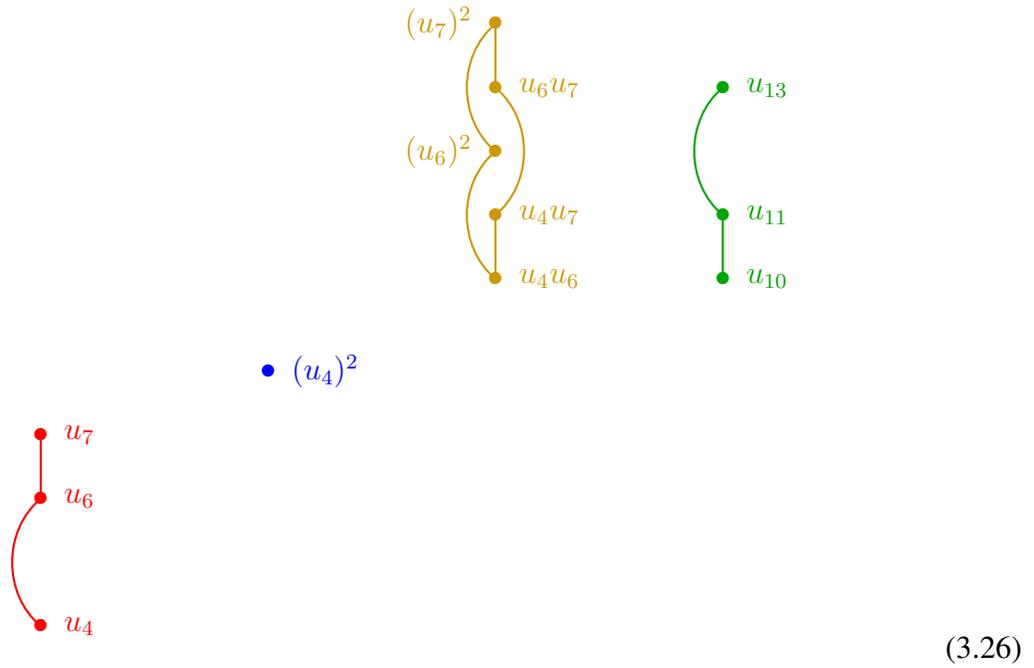
\begin{equation}
	\begin{tikzpicture}[thick]
		\matrix (m) [
			matrix of math nodes,
			row sep= 0.9em,
			column sep=6em
		]{
								&						& \colorC{\bullet}\\
								&						& \colorC{\bullet} & \colorE{\bullet}\\
								&						& \colorC{\bullet} & \\
								&						& \colorC{\bullet} & \colorE{\bullet}\\
								&						& \colorC{\bullet} & \colorE{\bullet}\\
								&						& \\
								& \colorG{\bullet} & \\
			\colorA{\bullet}\\
		    \colorA{\bullet}\\
			\phantom{\bullet}\\
			\colorA{\bullet} \\
		};
		\draw[colorA] (m-8-1.center) to (m-9-1.center);
		\draw[colorA] (m-9-1.center) to [out=220, in = 140] (m-11-1.center);
		\draw[colorC] (m-1-3.center) to (m-2-3.center);
		\draw[colorC] (m-4-3.center) to (m-5-3.center);
		\draw[colorC] (m-3-3.center) to [out=220, in = 140] (m-5-3.center);
		\draw[colorC] (m-1-3.center) to [out=220, in = 140] (m-3-3.center);
		\draw[colorC] (m-2-3.center) to [out=320, in = 40] (m-4-3.center);
		\draw[colorE] (m-4-4.center) to (m-5-4.center);
		\draw[colorE] (m-2-4.center) to [out=220, in = 140] (m-4-4.center);
		\node[colorA, anchor = west, inner sep=3mm] at (m-11-1) {$u_4$};
		\node[colorA, anchor = west, inner sep=3mm] at (m-9-1) {$u_6$};
		\node[colorA, anchor = west, inner sep=3mm] at (m-8-1) {$u_7$};
		\node[colorG, anchor = west, inner sep=3mm] at (m-7-2) {$(u_4)^2$};
		\node[colorC, anchor = west, inner sep=3mm] at (m-5-3) {$u_4u_6$};
		\node[colorC, anchor = west, inner sep=3mm] at (m-4-3) {$u_4u_7$};
		\node[colorC, anchor = east, inner sep=3mm] at (m-3-3) {$(u_{6})^2$};
		\node[colorC, anchor = east, inner sep=3mm] at (m-1-3) {$(u_{7})^2$};
		\node[colorC, anchor = west, inner sep=3mm] at (m-2-3) {$u_6u_7$};
		\node[colorE, anchor = west, inner sep=3mm] at (m-2-4) {$u_{13}$};
		\node[colorE, anchor = west, inner sep=3mm] at (m-4-4) {$u_{11}$};
		\node[colorE, anchor = west, inner sep=3mm] at (m-5-4) {$u_{10}$};
	\end{tikzpicture}
\end{equation}
which is namely
\begin{equation}
	\colorA{\Sigma^4 Q}
	\oplus
	\colorG{\Sigma^8 \bZ_2}
	\oplus
	\colorC{\Sigma^{10} J}
	\oplus
	\colorE{\Sigma^{10} \widetilde Q}
\end{equation}
where $\widetilde Q$ is a new (unnamed) module, and the corresponding Adams chart is given \cite{JohnFrancis2005} as
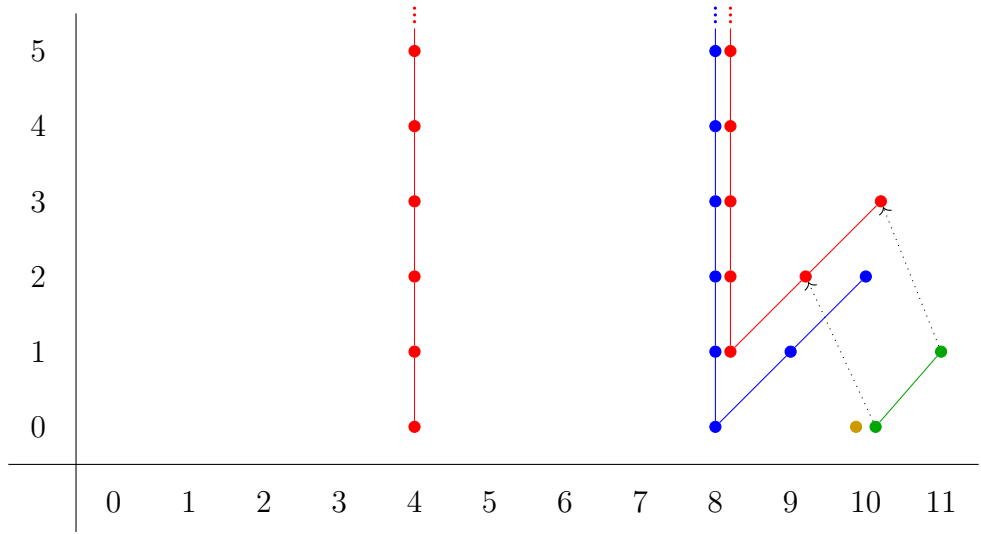
\begin{center}
\begin{sseqdata}[
	name=MKZ4,
	Adams grading,
	classes = fill,
	xrange = {0}{11},
	yrange = {0}{5}
]
	\tower[colorA](4,0)
	\tower[opacity=0](8,0)
	\tower[colorG](8,0)
	\class[opacity=0](8,0)
	\tower[colorA](8,1)
	\class[opacity=0](9,1)
	\class[colorG, opacity=1](9,1)
	\class[opacity=0](9,1)
	\structline[colorG](8,0,2)(9,1,2)
	\class[opacity=0](10,2)
	\class[colorG, opacity=1](10,2)
	\class[opacity=0](10,2)
	\structline[colorG](9,1,2)(10,2,2)
	\class[opacity=0](9,2)
	\class[opacity=0](9,2)
	\class[colorA,opacity=1](9,2)
	\structline[colorA](8,1,3)(9,2,3)
	\class[opacity=0](10,3)
	\class[opacity=0](10,3)
	\class[colorA,opacity=1](10,3)
	\structline[colorA](9,2,3)(10,3,3)
	\class[colorC, opacity=1](10,0)
	\class[colorE, opacity=1](10,0)
	\class[colorE, opacity=1](11,1)
	\structline[colorE](10,0,2)(11,1)
	\d[dotted]2 (11,1,,3)
	\d[dotted]2 (10,0,2,3)
\end{sseqdata}
\printpage[name = MKZ4, page=2]
\end{center}
According to \cite{JohnFrancis2005}, there are nontrivial differentials as shown by dotted lines,
and correspondingly the Adams spectral sequence converges as
\begin{equation}
	\label{omegaKZ4}
	\renewcommand{\arraystretch}{1.3}
	\begin{array}{c|ccccccccccccccccc}
		d & 0 & 1 & 2 & 3 & 4 & 5 & 6 & 7 & 8 & 9 & 10 & 11\\
		\hline
		\widetilde\Omega_{d}^{\text{spin}}(K(\bZ,4))
		& 0 & 0 & 0 & 0 & \bZ & 0 & 0 & 0 & \bZ^{\oplus 2} & \bZ_2 & \bZ_2^{\oplus 2} & 0\\
		\hline
	\end{array}
\end{equation}
In particular, the degree-9 part turns out to be
\begin{equation}
	\widetilde \Omega_9^{\text{spin}}(K(\bZ,4)) = \bZ_2
\end{equation}
which further describes the $E_{6,7,8}$ gauge anomaly based on the aforementioned reasoning.

As will be discussed in Sec.\,\ref{sec:GSmech},
this group also captures the anomaly of dynamical 2-form fields,
and as a result allows us to explain the cancellation of the universal gauge anomalies by the 2-form fields.

\subsection{$\mathrm{SO}(n)$ gauge anomaly}
The cohomology of $B\mathrm{SO}(n)$ is also known \cite{Borel1954, MimuraToda}
to be $p$-torsion free for $p\geq 3$,
so let us also look at the $G=\mathrm{SO}(n)$ case.
The $\bZ_2$ cohomology ring is well known and given as
\begin{equation}
	H^\ast(B\mathrm{SO}(n);\bZ_2) = \bZ_2[w_2, w_3, \ldots],
\end{equation}
where $w_i$'s are the Stiefel-Whitney classes, on which the cohomology operations act as
\begin{equation}
	\begin{array}{lclcl}
		Sq^1 w_i
		& = & (i-1)\,w_{i+1},\vspace{2mm}\\
		Sq^2 w_i
		& = & \left(\begin{array}{c}
			i-1\\
			2
		\end{array}\right) w_{i+2} + w_2 w_i.
	\end{array}
\end{equation}
Although $\mathrm{SO}(n)$ is not simply-connected,
the lowest degree of elements in $\widetilde H^\ast(B\mathrm{SO}(n);\bZ)$ is $2$,
meaning that one can derive the bordism group as in the previous case for $t-s \leq 9$,
which is barely sufficient for our purpose.
The $\cA(1)$-module structure of $H^\ast(B\mathrm{SO}(n);\bZ_2)$
for the range of interest (with large enough $n$) is represented as
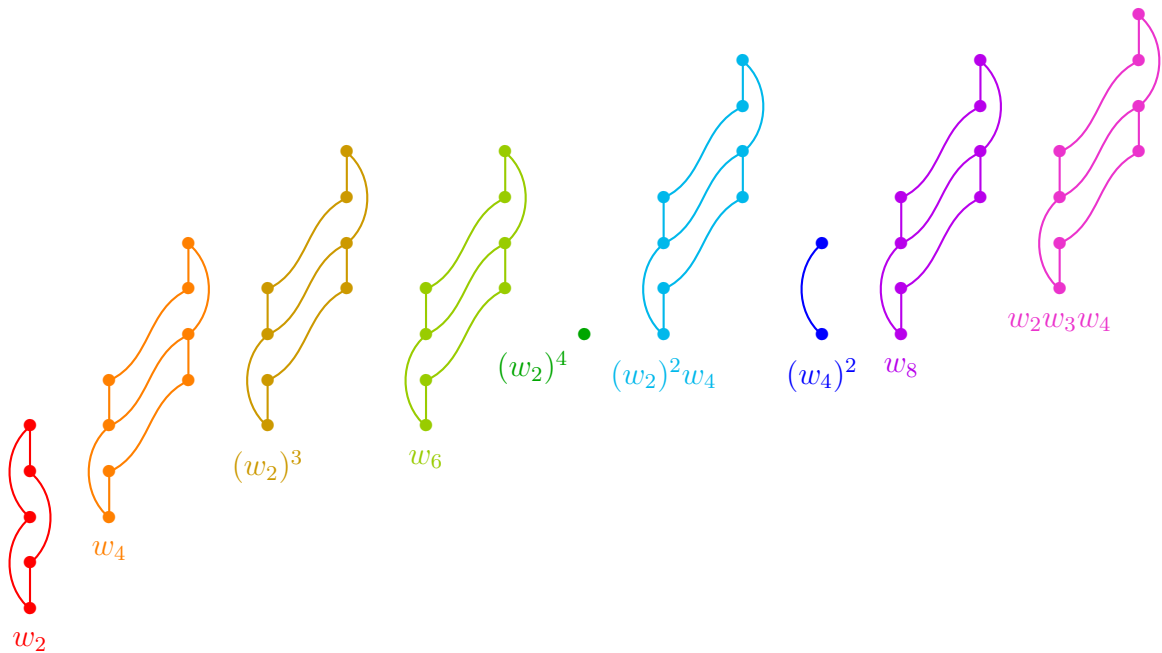
\begin{equation}
	\label{module-BSO}
	\begin{tikzpicture}[thick]
		\matrix (m) [
			matrix of math nodes,
			row sep= 0.33em,
			column sep=1.33em
		]{
									&&&&&&&&&						 &&&						  && \colorI{\bullet}\\
									&&&&&&&&& \colorF{\bullet} &&& \colorH{\bullet} && \colorI{\bullet}\\
									&&&&&&&&& \colorF{\bullet} &&& \colorH{\bullet} && \colorI{\bullet}\\
									&&&& \colorC{\bullet} && \colorD{\bullet} &&& \colorF{\bullet} &&& \colorH{\bullet} & \colorI{\bullet} & \colorI{\bullet}\\
									&							&							& 							& \colorC{\bullet}   &						 	 & \colorD{\bullet}	   &						 & \colorF{\bullet} & \colorF{\bullet} && \colorH{\bullet} & \colorH{\bullet} & \colorI{\bullet}\\
									&							& \colorB{\bullet}	 & 							 & \colorC{\bullet}   &							  & \colorD{\bullet}    &						  & \colorF{\bullet} && \colorG{\bullet} & \colorH{\bullet} && \colorI{\bullet}\\
									&							& \colorB{\bullet}	 & \colorC{\bullet}   & \colorC{\bullet}   & \colorD{\bullet}	& \colorD{\bullet}	  &							& \colorF{\bullet} &&& \colorH{\bullet} && \colorI{\bullet}\\
									&							& \colorB{\bullet}	 & \colorC{\bullet}	  &							  & \colorD{\bullet}   &							& \colorE{\bullet} & \colorF{\bullet}  && \colorG{\bullet} & \colorH{\bullet}\\
									& \colorB{\bullet}	 & \colorB{\bullet}	  & \colorC{\bullet}   &						   & \colorD{\bullet}\\
			\colorA{\bullet} & \colorB{\bullet}   &							  & \colorC{\bullet}   &						   & \colorD{\bullet}\\
			\colorA{\bullet} & \colorB{\bullet}\\
		    \colorA{\bullet} & \colorB{\bullet}\\
			\colorA{\bullet}\\
			\colorA{\bullet} \\
		};
		\draw[colorA] (m-10-1.center) to (m-11-1.center);
		\draw[colorA] (m-13-1.center) to (m-14-1.center);
		\draw[colorA] (m-12-1.center) to [out=220, in = 140] (m-14-1.center);
		\draw[colorA] (m-10-1.center) to [out=220, in = 140] (m-12-1.center);
		\draw[colorA] (m-11-1.center) to [out=320, in = 40] (m-13-1.center);
		\draw[colorB] (m-9-2.center) to (m-10-2.center);
		\draw[colorB] (m-11-2.center) to (m-12-2.center);
		\draw[colorB] (m-10-2.center) to [out=220, in = 140] (m-12-2.center);
		\draw[colorB] (m-11-2.center) to [out=25, in = 205] (m-9-3.center);
		\draw[colorB] (m-10-2.center) to [out=25, in = 205] (m-8-3.center);
		\draw[colorB] (m-9-2.center) to [out=25, in = 205] (m-7-3.center);
		\draw[colorB] (m-8-3.center) to (m-9-3.center);
		\draw[colorB] (m-6-3.center) to (m-7-3.center);
		\draw[colorB] (m-6-3.center) to [out=320, in = 40] (m-8-3.center);
		\draw[colorC] (m-7-4.center) to (m-8-4.center);
		\draw[colorC] (m-9-4.center) to (m-10-4.center);
		\draw[colorC] (m-8-4.center) to [out=220, in = 140] (m-10-4.center);
		\draw[colorC] (m-9-4.center) to [out=25, in=205] (m-7-5.center);
		\draw[colorC] (m-8-4.center) to [out=25, in=205] (m-6-5.center);
		\draw[colorC] (m-7-4.center) to [out=25, in=205] (m-5-5.center);
		\draw[colorC] (m-6-5.center) to (m-7-5.center);
		\draw[colorC] (m-4-5.center) to (m-5-5.center);
		\draw[colorC] (m-4-5.center) to [out=320, in = 40] (m-6-5.center);
		\draw[colorD] (m-7-6.center) to (m-8-6.center);
		\draw[colorD] (m-9-6.center) to (m-10-6.center);
		\draw[colorD] (m-8-6.center) to [out=220, in = 140] (m-10-6.center);
		\draw[colorD] (m-9-6.center) to [out=25, in=205] (m-7-7.center);
		\draw[colorD] (m-8-6.center) to [out=25, in=205] (m-6-7.center);
		\draw[colorD] (m-7-6.center) to [out=25, in=205] (m-5-7.center);
		\draw[colorD] (m-6-7.center) to (m-7-7.center);
		\draw[colorD] (m-4-7.center) to (m-5-7.center);
		\draw[colorD] (m-4-7.center) to [out=320, in = 40] (m-6-7.center);
		\draw[colorF] (m-5-9.center) to (m-6-9.center);
		\draw[colorF] (m-7-9.center) to (m-8-9.center);
		\draw[colorF] (m-6-9.center) to [out=220, in = 140] (m-8-9.center);
		\draw[colorF] (m-7-9.center) to [out=25, in=205] (m-5-10.center);
		\draw[colorF] (m-6-9.center) to [out=25, in=205] (m-4-10.center);
		\draw[colorF] (m-5-9.center) to [out=25, in=205] (m-3-10.center);
		\draw[colorF] (m-4-10.center) to (m-5-10.center);
		\draw[colorF] (m-2-10.center) to (m-3-10.center);
		\draw[colorF] (m-2-10.center) to [out=320, in = 40] (m-4-10.center);
		\draw[colorG] (m-6-11.center) to [out=220, in = 140] (m-8-11.center);
		\draw[colorH] (m-5-12.center) to (m-6-12.center);
		\draw[colorH] (m-7-12.center) to (m-8-12.center);
		\draw[colorH] (m-6-12.center) to [out=220, in = 140] (m-8-12.center);
		\draw[colorH] (m-7-12.center) to [out=25, in=205] (m-5-13.center);
		\draw[colorH] (m-6-12.center) to [out=25, in=205] (m-4-13.center);
		\draw[colorH] (m-5-12.center) to [out=25, in=205] (m-3-13.center);
		\draw[colorH] (m-4-13.center) to (m-5-13.center);
		\draw[colorH] (m-2-13.center) to (m-3-13.center);
		\draw[colorH] (m-2-13.center) to [out=320, in = 40] (m-4-13.center);
		\draw[colorI] (m-4-14.center) to (m-5-14.center);
		\draw[colorI] (m-6-14.center) to (m-7-14.center);
		\draw[colorI] (m-5-14.center) to [out=220, in = 140] (m-7-14.center);
		\draw[colorI] (m-6-14.center) to [out=25, in=205] (m-4-15.center);
		\draw[colorI] (m-5-14.center) to [out=25, in=205] (m-3-15.center);
		\draw[colorI] (m-4-14.center) to [out=25, in=205] (m-2-15.center);
		\draw[colorI] (m-3-15.center) to (m-4-15.center);
		\draw[colorI] (m-1-15.center) to (m-2-15.center);
		\draw[colorI] (m-1-15.center) to [out=320, in = 40] (m-3-15.center);
		\node[colorA, anchor = north, inner sep=3mm] at (m-14-1) {$w_2$};
		\node[colorB, anchor = north, inner sep=3mm] at (m-12-2) {$w_4$};
		\node[colorC, anchor = north, inner sep=3mm] at (m-10-4) {$(w_2)^3$};
		\node[colorD, anchor = north, inner sep=3mm] at (m-10-6) {$w_6$};
		\node[colorE, anchor = north east, inner sep=2mm] at (m-8-8) {$(w_2)^4$};
		\node[colorF, anchor = north, inner sep=3mm] at (m-8-9) {$(w_2)^2w_4$};
		\node[colorG, anchor = north, inner sep=3mm] at (m-8-11) {$(w_4)^2$};
		\node[colorH, anchor = north, inner sep=3mm] at (m-8-12) {$w_8$};
		\node[colorI, anchor = north, inner sep=3mm] at (m-7-14) {$w_2w_3w_4$};
	\end{tikzpicture}
\end{equation}
which is namely
\begin{equation}
	\colorA{\Sigma^2 J}
	\oplus
	\colorB{\Sigma^4 \cA(1)}
	\oplus
	\colorC{\Sigma^6 \cA(1)}
	\oplus
	\colorD{\Sigma^6 \cA(1)}
	\oplus
	\colorE{\Sigma^8 \bZ_2}
	\oplus
	\colorF{\Sigma^8 \cA(1)}
	\oplus
	\colorG{\Sigma^8 \cA(1)/\!\!/ \cE(1)}
	\oplus
	\colorH{\Sigma^8 \cA(1)}
	\oplus
	\colorI{\Sigma^9 \cA(1)}
\end{equation}
and the corresponding Adams chart is
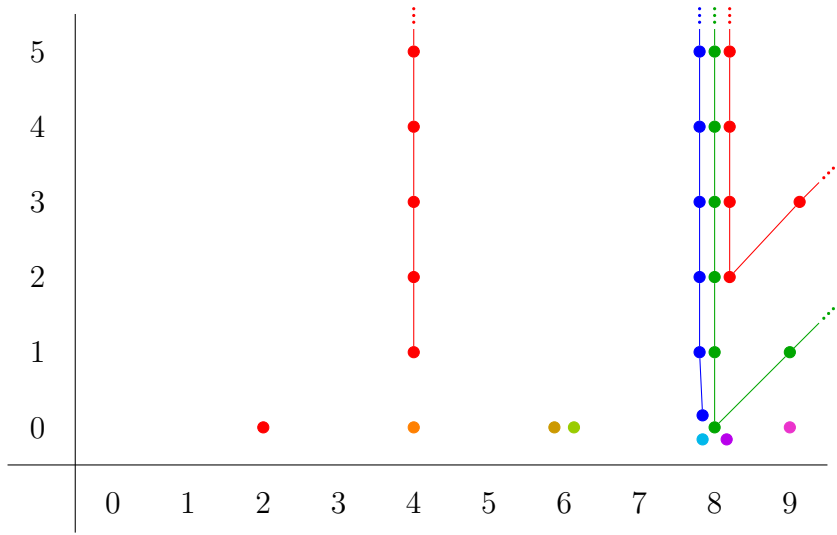
\begin{center}
\begin{sseqdata}[
	name=MBSO,
	Adams grading,
	classes = fill,
	xrange = {0}{9},
	yrange = {0}{5}
]
	\class[colorA](2,0)
	\tower[colorA](4,1)
	\class[colorB](4,0)
	\class[colorC](6,0)
	\class[colorD](6,0)
	\class[colorF](8,0)
	\class[colorH](8,0)
	\class[colorI](9,0)
	\tower[colorG](8,0)
	\class[opacity=0] (8,0)
	\tower[colorE](8,0)
	\class[colorE] (9,1)
	\class[opacity=0] (10,2)
	\structline[colorE](8,0,5)(9,1,1)
	\structline[colorE](9,1,1)(10,2,1)
	\tower[colorA](8,2)
	\tower[opacity=0] (9,3)
	\class[colorA](9,3)
	\structline[colorA](8,2,3)(9,3,2)
	\class[opacity=0] (10,4)
	\class[opacity=0] (10,4)
	\structline[colorA](9,3,2)(10,4,2)
	\class[opacity=0] (8,1)
\end{sseqdata}
\printpage[name = MBSO, page=2]
\end{center}
As before, there are no differentials from $t-s \leq 9$, and the Adams spectral sequence converges as
\begin{equation}
	\label{omegaBSO}
	\renewcommand{\arraystretch}{1.3}
	\begin{array}{c|ccccccccccccccccc}
		d & 0 & 1 & 2 & 3 & 4 & 5 & 6 & 7 & 8 & \cdots \\
		\hline
		\widetilde\Omega_{d}^{\text{spin}}(B\mathrm{SO}(n))
		& 0 & 0 & \bZ_2 & 0 & \bZ\oplus \bZ_2 & 0 & \bZ_2^{\oplus 2} & 0 & \bZ^{\oplus 3}\oplus \bZ_2^{\oplus 2} & \cdots\\
		\hline
	\end{array}
\end{equation}
and the degree-9 part contains at least two $\bZ_2$'s corresponding to $(t-s,s) = (9,0)$ and $(9,1)$
which cannot be killed by differentials from $t-s=10$.

\subsection{$\mathrm{Spin}(n)$ gauge anomaly}
Also, the $\bZ_2$ cohomology of $B\mathrm{Spin}(n)$ is known \cite[Theorem\,6.5]{QuillenBSpin} to be
\begin{equation}
	H^\ast(B\mathrm{Spin}(n); \bZ_2)
	\simeq
	H^\ast(B\mathrm{SO}(n); \bZ_2)/J
	\otimes
	\bZ_2[w_{2^h}(\Delta_\theta)]
\end{equation}
where $h\approx n/2$ and $J$ is an ideal generated by
\begin{equation}
	\renewcommand{\arraystretch}{1.3}
	\begin{array}{r}
		w_2,\\
		Sq^1w_2,\\
		\vdots \hspace*{12pt}\\
		Sq^{2^{h-1}}Sq^{2^{h-2}}\cdots Sq^{1}w_2
	\end{array}
\end{equation}
and therefore effectively removes $w_2$, $w_3$, $w_5$, $w_9$, ... from the \eqref{module-BSO}, resulting in
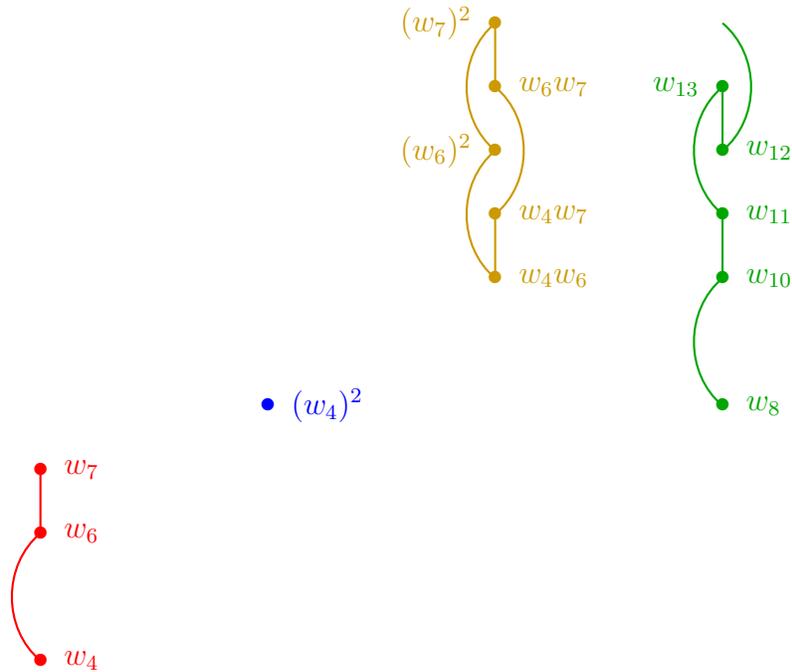
\begin{equation}
	\begin{tikzpicture}[thick]
		\matrix (m) [
			matrix of math nodes,
			row sep= 0.9em,
			column sep=6em
		]{
								&						& \colorC{\bullet} & \textcolor{white}{\bullet}\\
								&						& \colorC{\bullet} & \colorE{\bullet}\\
								&						& \colorC{\bullet} & \colorE{\bullet}\\
								&						& \colorC{\bullet} & \colorE{\bullet}\\
								&						& \colorC{\bullet} & \colorE{\bullet}\\
								&						& 						  & \phantom{\bullet}\\
								& \colorG{\bullet} & 						& \colorE{\bullet}\\
			\colorA{\bullet}\\
		    \colorA{\bullet}\\
			\phantom{\bullet}\\
			\colorA{\bullet} \\
		};
		\draw[colorA] (m-8-1.center) to (m-9-1.center);
		\draw[colorA] (m-9-1.center) to [out=220, in = 140] (m-11-1.center);
		\draw[colorC] (m-1-3.center) to (m-2-3.center);
		\draw[colorC] (m-4-3.center) to (m-5-3.center);
		\draw[colorC] (m-3-3.center) to [out=220, in = 140] (m-5-3.center);
		\draw[colorC] (m-1-3.center) to [out=220, in = 140] (m-3-3.center);
		\draw[colorC] (m-2-3.center) to [out=320, in = 40] (m-4-3.center);
		\draw[colorE] (m-2-4.center) to (m-3-4.center);
		\draw[colorE] (m-4-4.center) to (m-5-4.center);
		\draw[colorE] (m-1-4.center) to [out=320, in = 40] (m-3-4.center);
		\draw[colorE] (m-2-4.center) to [out=220, in = 140] (m-4-4.center);
		\draw[colorE] (m-5-4.center) to [out=220, in = 140] (m-7-4.center);
		\node[colorA, anchor = west, inner sep=3mm] at (m-11-1) {$w_4$};
		\node[colorA, anchor = west, inner sep=3mm] at (m-9-1) {$w_6$};
		\node[colorA, anchor = west, inner sep=3mm] at (m-8-1) {$w_7$};
		\node[colorG, anchor = west, inner sep=3mm] at (m-7-2) {$(w_4)^2$};
		\node[colorC, anchor = west, inner sep=3mm] at (m-5-3) {$w_4w_6$};
		\node[colorC, anchor = west, inner sep=3mm] at (m-4-3) {$w_4w_7$};
		\node[colorC, anchor = east, inner sep=3mm] at (m-3-3) {$(w_{6})^2$};
		\node[colorC, anchor = east, inner sep=3mm] at (m-1-3) {$(w_{7})^2$};
		\node[colorC, anchor = west, inner sep=3mm] at (m-2-3) {$w_6w_7$};
		\node[colorE, anchor = east, inner sep=3mm] at (m-2-4) {$w_{13}$};
		\node[colorE, anchor = west, inner sep=3mm] at (m-4-4) {$w_{11}$};
		\node[colorE, anchor = west, inner sep=3mm] at (m-5-4) {$w_{10}$};
		\node[colorE, anchor = west, inner sep=3mm] at (m-7-4) {$w_{8}$};
		\node[colorE, anchor = west, inner sep=3mm] at (m-3-4) {$w_{12}$};
	\end{tikzpicture}
\end{equation}
for large enough $n$.
The corresponding Adams chart is (cf.~\cite{JohnFrancis2005})
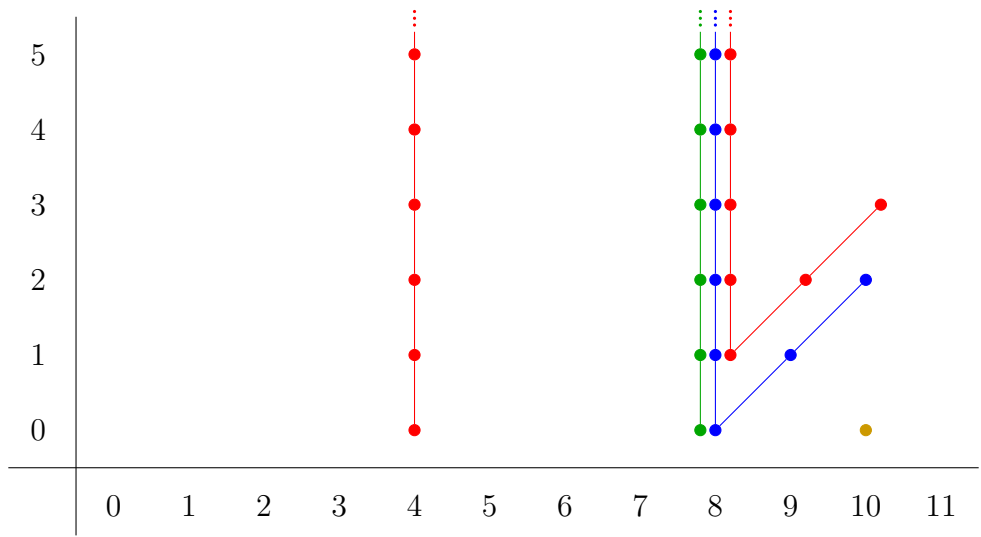
\begin{center}
\begin{sseqdata}[
	name=MBSpin,
	Adams grading,
	classes = fill,
	xrange = {0}{11},
	yrange = {0}{5}
]
	\tower[colorA](4,0)
	\tower[colorE, opacity=1](8,0)
	\tower[colorG](8,0)
	\class[opacity=0](8,0)
	\tower[colorA](8,1)
	\class[opacity=0](9,1)
	\class[colorG, opacity=1](9,1)
	\class[opacity=0](9,1)
	\structline[colorG](8,0,2)(9,1,2)
	\class[opacity=0](10,2)
	\class[colorG, opacity=1](10,2)
	\class[opacity=0](10,2)
	\structline[colorG](9,1,2)(10,2,2)
	\class[opacity=0](9,2)
	\class[opacity=0](9,2)
	\class[colorA,opacity=1](9,2)
	\structline[colorA](8,1,3)(9,2,3)
	\class[opacity=0](10,3)
	\class[opacity=0](10,3)
	\class[colorA,opacity=1](10,3)
	\structline[colorA](9,2,3)(10,3,3)
	\class[colorC, opacity=1](10,0)
\end{sseqdata}
\printpage[name = MBSpin, page=2]
\end{center}
with no differentials at all.
As a result one obtains
\begin{equation}
	\label{omegaBSpin}
	\renewcommand{\arraystretch}{1.3}
	\begin{array}{c|ccccccccccccccccc}
		d & 0 & 1 & 2 & 3 & 4 & 5 & 6 & 7 & 8 & 9 & 10 & 11\\
		\hline
		\widetilde\Omega_{d}^{\text{spin}}(B\mathrm{Spin}(n))
		& 0 & 0 & 0 & 0 & \bZ & 0 & 0 & 0 & \bZ^{\oplus 3} & \bZ_2^{\oplus 2} & \bZ_2^{\oplus 3} & 0\\
		\hline
	\end{array}
\end{equation}

\bigskip

\subsection{$\bZ_2$ 3-form fields}\label{sec:Z2}
Here we also compute the bordism group for $X=K(\bZ_2, 4)$ for later use in Sec.\,\ref{sec:TQFT}.
It is supposed to capture the anomalies of 3-form $\bZ_2$ gauge fields,
in a similar vein to the 2-form fields' case.

The $\bZ_2$ cohomology of the Eilenberg-MacLane space $K(\bZ_2, 4)$ is known \cite{Serre1953, MimuraToda} to be
\begin{equation}
	\setlength{\arraycolsep}{3pt}
	\begin{array}{lllll}
		H^\ast(K(\bZ_2,4);\bZ_2) = \bZ_2[ & u_4,\\
		& Sq^1u_4, & Sq^2u_4, & Sq^3u_4,\\
		& Sq^2Sq^1 u_4, & Sq^3Sq^1u_4, & Sq^4Sq^1u_4,\\
		& Sq^4Sq^2 u_4, & Sq^5Sq^2 u_4, & Sq^6Sq^3 u_4, & Sq^4Sq^2Sq^1 u_4, ...],
	\end{array}
\end{equation}
where the generators with more than two Steenrod squares irrelevant to our purpose are omitted.
The $\cA(1)$-module structure of $H^\ast(K(\bZ_2,4);\bZ_2)$ for the range of interest is represented as
\begin{equation}
	\begin{tikzpicture}[thick]
		\matrix (m) [
			matrix of math nodes,
			row sep= 0.9em,
			column sep=3em
		]{
									&						  &&&							   &						 &&						 	& \colorF{\bullet}\\
									&						  &&&							   &						 &&						 	& \colorF{\bullet}\\
									&						  &&& \colorC{\bullet}		&						  &&						 & \colorF{\bullet}\\
									&						  &&& \phantom{\bullet}  & \colorE{\bullet}  && \colorF{\bullet} & \colorF{\bullet}\\
									&						  &&& \colorC{\bullet}		&						  && \colorF{\bullet}\\
									&						  &&& \colorC{\bullet}		& \colorE{\bullet} && \colorF{\bullet}\\
									& \colorA{\bullet} &&&								& \colorE{\bullet} && \colorF{\bullet}\\
									& \colorA{\bullet} &&& \colorC{\bullet}\\
									& \colorA{\bullet} && \colorG{\bullet}\\
			\colorA{\bullet} & \colorA{\bullet} \\
		    \colorA{\bullet}\\
			\colorA{\bullet}\\
			\colorA{\bullet} \\
		};
		\draw[colorA] (m-10-1.center) to (m-11-1.center);
		\draw[colorA] (m-12-1.center) to (m-13-1.center);
		\draw[colorA] (m-11-1.center) to [out=220, in = 140] (m-13-1.center);
		\draw[colorA] (m-12-1.center) to [out=25, in = 195] (m-10-2.center);
		\draw[colorA] (m-11-1.center) to [out=25, in = 195] (m-9-2.center);
		\draw[colorA] (m-10-1.center) to [out=25, in = 195] (m-8-2.center);
		\draw[colorA] (m-7-2.center) to (m-8-2.center);
		\draw[colorA] (m-9-2.center) to (m-10-2.center);
		\draw[colorA] (m-9-2.center) to [out=40, in = 320] (m-7-2.center);
		\node[colorA, anchor = west, inner sep=3mm] at (m-13-1) {$u_4$};
		\node[colorA, anchor = east, inner sep=6mm] at (m-12-1) {$Sq^1u_4$};
		\node[colorA, anchor = east, inner sep=6mm] at (m-11-1) {$Sq^2u_4$};
		\node[colorA, anchor = east, inner sep=6mm] at (m-10-1) {$Sq^3u_4$};
		\node[colorA, anchor = west, inner sep=6mm] at (m-10-2) {$Sq^2Sq^1u_4$};
		\node[colorA, anchor = west, inner sep=6mm] at (m-9-2) {$Sq^3Sq^1u_4$};
		\node[colorA, anchor = west, inner sep=6mm] at (m-8-2) {$Sq^4Sq^1u_4$};
		\node[colorA, anchor = west, inner sep=6mm] at (m-7-2) {$(Sq^1u_4)^2$};
		\node[colorG, anchor = north, inner sep=3mm] at (m-9-4) {\phantom{aaaaaaaaaaaaa}$(u_4)^2 + Sq^3Sq^1u_4$};
		\node[colorC, anchor = north west, inner sep=2mm] at (m-8-5) {$u_4 Sq^1 u_4 + Sq^4Sq^1u_4$};
		\draw[colorC] (m-6-5.center) to [out=220, in = 140] (m-8-5.center);
		\draw[colorC] (m-6-5.center) to (m-5-5.center);
		\draw[colorC] (m-3-5.center) to [out=220, in = 140] (m-5-5.center);
		\node[colorE, anchor = west, inner sep=3mm] at (m-7-6) {$Sq^4 Sq^2 u_4$};
		\node[colorE, anchor = west, inner sep=3mm] at (m-6-6) {$Sq^5 Sq^2 u_4$};
		\node[colorE, anchor = west, inner sep=3mm] at (m-4-6) {$Sq^6 Sq^3 u_4$};
		\draw[colorE] (m-4-6.center) to [out=220, in = 140] (m-6-6.center);
		\draw[colorE] (m-6-6.center) to (m-7-6.center);
		\node[colorF, anchor = west, inner sep=3mm] at (m-7-8) {$u_4 Sq^2 u_4$};
		\draw[colorF] (m-4-8.center) to (m-5-8.center);
		\draw[colorF] (m-6-8.center) to (m-7-8.center);
		\draw[colorF] (m-5-8.center) to [out=220, in = 140] (m-7-8.center);
		\draw[colorF] (m-6-8.center) to [out=25, in = 195] (m-4-9.center);
		\draw[colorF] (m-5-8.center) to [out=25, in = 195] (m-3-9.center);
		\draw[colorF] (m-4-8.center) to [out=25, in = 195] (m-2-9.center);
		\draw[colorF] (m-1-9.center) to (m-2-9.center);
		\draw[colorF] (m-3-9.center) to (m-4-9.center);
		\draw[colorF] (m-3-9.center) to [out=40, in = 320] (m-1-9.center);
	\end{tikzpicture}
\end{equation}
which is namely
\begin{equation}
	\colorA{\Sigma^4 \cA(1)}
	\oplus \colorG{\Sigma^8 \bZ_2}
	\oplus \colorC{\Sigma^9 \cA(1) /\!\!/ \cE(0)}
	\oplus \colorE{\Sigma^{10} \widetilde Q}
	\oplus \colorF{\Sigma^{10} \cA(1)}
\end{equation}
and the corresponding Adams chart is
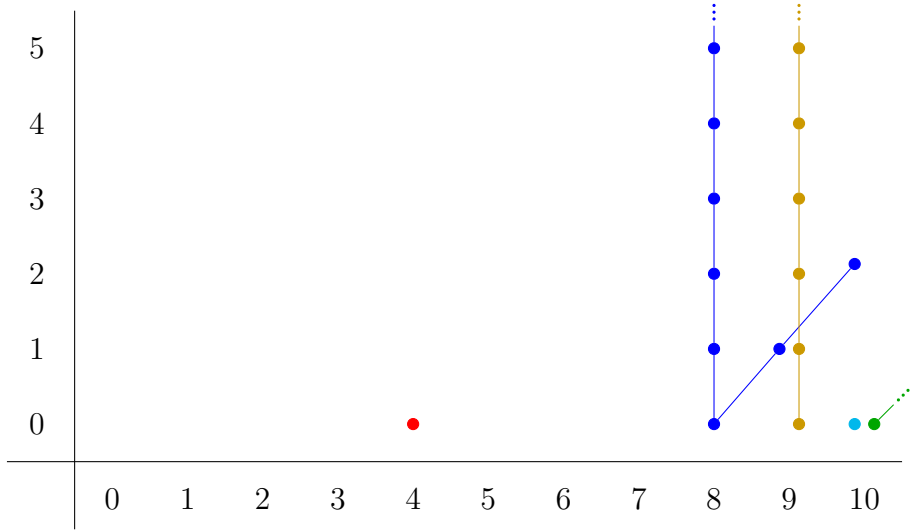
\begin{center}
\begin{sseqdata}[
	name=MKZ2-4,
	Adams grading,
	classes = fill,
	xrange = {0}{10},
	yrange = {0}{5}
]
	\class[colorA](4,0)
	\tower[colorG](8,0)
	\class[colorG](9,1)
	\class[opacity=0] (10,2)
	\class[opacity=0] (10,2)
	\class[colorG] (10,2)
	\class[opacity=0](10,2)
	\structline[colorG](8,0)(9,1)
	\structline[colorG](9,1,1)(10,2,3)
	\class[opacity=0] (9,0)
	\class[opacity=0] (9,2)
	\class[opacity=0] (9,3)
	\class[opacity=0] (9,4)
	\class[opacity=0] (9,5)
	\class[opacity=0] (9,6)
	\tower[colorC](9,0)	
	\class[colorF] (10,0)
	\class[colorE] (10,0)
	\class[opacity=0] (11,1)
	\class[colorE] (11,1)
	\structline[colorE](10,0,2)(11,1,2)
\end{sseqdata}
\printpage[name = MKZ2-4, page=2]
\end{center}
Although the (part of) towers at $t-s=9$ inevitably interfere with that at $t-s=8$,
there is no differential which can kill the other $(t-s,s)=(9,1)$ element, and it is guaranteed to survive.
Therefore, one can deduce
\begin{equation}
	\widetilde\Omega_9^{\text{spin}}(K(\bZ_2, 4))_2^\wedge \supset \bZ_2
\end{equation}
and it is inferred that $\bZ_2$ 3-form fields carry (at least) an order-2 anomaly.

\subsection{Structure of generator manifolds}\label{sec:str}
The sloped lines in Adams charts representing $h_1\in \mathrm{Ext}_{\cA}^{1,2}(\bZ_2, \bZ_2)$
correspond to multiplication $ \pi^\mathrm{st}_1(\mathrm{pt}) \times  \pi^\textrm{st}_\bullet(-) \to  \pi^\textrm{st}_{\bullet+1}(-)$
in terms of stable homotopy groups \cite{Hatcher}.
Under the Pontrjagin-Thom construction, this multiplication can be geometrically interpreted as $[S^1]\times [M] \to [ S^1 \times M]$,
where $M$ is a manifold representing an element $[M] \in \widetilde \Omega_\bullet^{\text{spin}}(X)$, and also $S^1$ is a representative manifold of the 
nontrivial element of $ \pi^\textrm{st}_1(\mathrm{pt})  \simeq  \pi^\textrm{st}_1(M\mathrm{Spin}) = \Omega_1^{\text{spin}}(\mathrm{pt}) = \bZ_2$.
In particular, the elements of $\widetilde \Omega_9^{\text{spin}}(X) $ obtained in this section which stem from elements of $\widetilde \Omega_8^{\text{spin}}(X)$
by a sloped line are represented as $[S^1 \times M_8]$ for some $[M_8] \in \widetilde \Omega_8^{\text{spin}}(X)$,
and this is in fact how we obtained the examples of representative manifolds in Sec.\,\ref{sec:explicit}.

Moreover, for the elements $[M_t] \in \widetilde \Omega_{t-0}^{\text{spin}}(X)$ coming from Adams filtration $s=0$,
i.e. $E_2^{s=0,t} = \mathrm{Ext}_{\cA(1)}^{s=0,t}\big(\widetilde H^\ast( X;\bZ_2), \bZ_2) =  
\mathrm{Hom}_{\cA(1)}^{t}\big(\widetilde H^\ast( X;\bZ_2), \bZ_2)$,
there may be a simple interpretation in terms of cohomology (see e.g.~\cite{Freed:2019sco}). 
Let $f: M_t \to X$ be a representative of an element of $ \widetilde \Omega_t^{\text{spin}}(X)$
and let us label an element in the row $s=0$ by a cohomology class $c_t \in H^t(X;\bZ_2)$.
Then the integral $\int_{M_t} f^* c_t \in \bZ_2$ (or more precisely the evaluation of $f^* c_t$ by the fundamental class of $M_t$) has a nontrivial value.

As an example, consider the case $X = K(\bZ,4)$.
We have the element $u_4 \in H^4(K(\bZ,4); \bZ_2)$. Taking a map $f : \mathbb{HP}^2 \to K(\bZ,4)$
such that the pullback $f^* u_4$ is (a $\bZ_2$ reduction of) $x \in H^4(\mathbb{HP}^2;\bZ)$,
the cohomology class $(u_4)^2$ has a nontrivial value as $\int_{\mathbb{HP}^2} f^* (u_4)^2 = 1 \in \bZ_2$.
In this way, we see that this $f : \mathbb{HP}^2 \to K(\bZ,4)$ represents a nontrivial element of $ \widetilde \Omega_8^{\text{spin}}(K(\bZ,4))$,
which can be detected by the cohomology class $(u_4)^2 \in H^8(K(\bZ,4); \bZ_2)$. 

Similarly, in the case of $X=K(\bZ_2,4)$, we have seen that the cohomology class labelling the nontrivial element of $ \widetilde \Omega_8^{\text{spin}}(K(\bZ_2,4))$
is $(u_4)^2 + Sq^3 Sq^1u_4$, but since $f^\ast(Sq^1 u_4)=0$ on $\mathbb{HP}^2$, the argument reduces to that of $K(\bZ,4)$.
The nontrivial element of $ \widetilde \Omega_9^{\text{spin}}(K(\bZ,4)) = \bZ_2$ (or the analogous element of $ \widetilde \Omega_9^{\text{spin}}(K(\bZ_2,4))$) 
is simply obtained by multiplying $S^1$ as discussed above.

\section{Anomaly cancellation via 2-form fields}\label{sec:GSmech}
In the previous sections, we have found that a fermion in the adjoint representation
always has an anomaly for any simple Lie group $G$, 
which is detected by $G$-bundles $P_G \to \mathbb{HP}^2$ ($\times S^1$). 
Also, a gravitino had a pure gravitational anomaly detected by $\mathbb{HP}^2$ ($\times S^1$) which cannot be cancelled by spin $1/2$ fermions.
However, we know that both an adjoint fermion
(namely gaugino) and a gravitino are realized in string theory with $\cN=1$ supersymmetry in 8-dimensions for 
$G = \SU(n), \Spin(2n), \Sp(n), E_{6,7,8}$ (e.g. by F-theory),
so there must be a mechanism to cancel these anomalies. 
In this section, we discuss anomaly cancellation via 2-form fields,
which is exactly a non-perturbative version of the Green-Schwarz mechanism.

\subsection{2-form fields}
A dynamical 2-form field $B$ in $d$ spacetime dimensions
yields two conserved currents $j_e \sim \ast dB$ (where $\ast$ is the Hodge star) and $j_m \sim dB$, and correspondingly the theory actually possesses
electric 2-form $\mathrm{U}(1)$ symmetry and 
magnetic $(d-4)$-form $\mathrm{U}(1)$ symmetry.
Modern understanding of the Green-Schwarz mechanism (and its relatives) is that, it should be interpreted as describing
a 't Hooft anomaly of these higher-form symmetries~\cite{Gaiotto:2014kfa},
which enables the cancellation against other anomalies after turning on their background gauge fields $A_e$ and $A_m$ \cite{Hsieh:2020jpj}.

Our claim is that
the global anomaly of the theory when it is coupled to the 3-form field  $A_e$ corresponds to 
an element of $\Hom(\widetilde \Omega^{\text{spin}}_{d+1}(K(\bZ,4)) , \U(1))$.\footnote{
	For more general case with nonzero perturbative anomaly, the anomalies should correspond to elements of the Anderson dual $(\widetilde{I\Omega}{}^{\text{spin}})^{d+2}(K(\bZ,4) )$ of the bordism group \cite{Freed:2016rqq, Yamashita:2021cao}. 
} Here, $K(\bZ,4)$ appears because the topology of the background 3-form field $A_e$ is classified by its 4-form flux (or more precisely its integral-cohomology version).
For our purpose, $A_e$ will be taken to be a Chern-Simons 3-form of the $G$ gauge field. 
To explain the anomaly, we construct a $(d+1)$-dimensional bulk theory
which hosts the original theory on the boundary.
We follow the discussions in \cite{Hsieh:2020jpj} suitably modified according to the present situation.

\bigskip

Let 
$Q$ be an action in $(d+1$)-dimensions describing the anomaly in $d$-dimensions in question.
For the purpose of this paper, we are merely concerned with global anomalies and thus $Q$ is taken to be
an element of $\Hom(\widetilde \Omega^{\text{spin}}_{d+1}(K(\bZ,4)) , \mathrm{U}(1))$,
but we remark that the discussions below can in principle be generalized to the case where perturbative anomalies are present,
especially the case of the original $10d$ Green-Schwarz mechanism.\footnote{
We leave it for future work to describe the details of the $10d$ case.} 

Let us introduce a dynamical 3-form field $C$ and a dynamical $(d-3)$-form field $D$, both in $(d+1)$-dimensional bulk.\footnote{
	This can be thought of as an analog of the realization of chiral fermions as boundary modes of massive fermions in one-higher dimensions;
	the dynamical 2-form field $B$ in question corresponds to a chiral fermion,
	which is to be realized as a boundary mode of a ``massive'' dynamical 3-form field $C$.
}
All the $p$-form fields are normalized so that their fluxes are integer-valued. Then, we take the Euclidean action given by
\beq
S = -\frac{1}{2} \int_{W_{d+1}}\left( \frac{1}{e^{2}}\, dC \wedge * dC + \frac{1}{e'^{2}}\, dD \wedge *dD \right) + 2\pi i \int_{W_{d+1}} D \wedge d (C -A_e) + 2\pi i\cdot  Q(C) \label{eq:bulkaction}
\eeq
where $e$ and $e'$ are parameters, and $*$ is the Hodge star. The product $ee'$ has mass dimension 1.
More precise definitions of ``$p$-form fields'' and terms like ``$\int D \wedge dC$\,'' are given by the theory of differential cohomology~\cite{CS}.\footnote{
	See e.g. \cite{Freed:2006yc,Hsieh:2020jpj} for reviews aimed at physicists.
} 

First, let us consider the above theory on a closed $(d+1)$-manifold $W_{d+1}$ (i.e. 
$\partial W_{d+1} = \varnothing$).
After taking the limit $e, e' \to \infty$, the kinetic term can be neglected.
Carrying out the path integral over $D$ which serves as a Lagrange multiplier setting $C \to A_e$, we get
\beq
S \to   2\pi i \cdot Q(A_e).
\eeq
In this way, the bulk theory only depends on the background field $A_e$, and does not have any dynamical degrees of freedom.

Next, let us put the theory on a manifold $W_{d+1}$ with boundary $\partial W_{d+1} = M_d$.
Here we impose a standard Dirichlet-type boundary condition such that 
\beq
C|_{\partial W_{d+1}} =0, \qquad D|_{\partial W_{d+1}} =0. \label{eq:boundarycondition}
\eeq 
Under this boundary condition, the second and third terms of \eqref{eq:bulkaction} 
indeed make sense 
for the following reason~\cite{Hsieh:2020jpj}. Take any manifold $W_{d+1}'$ 
with the same boundary $M_d$ but with the opposite orientation to $W_{d+1}$, so that we can glue them to 
get a closed manifold \mbox{$W_{\text{closed}} = W_{d+1} \cup W_{d+1}'$}.
Since $C$ and $D$ vanish on the boundary $M_d = \partial W_{d+1}$, we can trivially extend them by demanding that they are zero on $W_{d+1}'$.
In this way, we get field configurations on the entire manifold $W_{\text{closed}}$. 
Then, we define the values for $2\pi i \int D \wedge d (C -A) $ and $2\pi i \cdot Q(C)$ on $W_{d+1}$ 
to be those on $W_{\text{closed}}$, which can be safely obtained. 
These values do not depend on the choice of $W_{d+1}'$; 
the possible difference 
between two choices $W_{d+1}'$ and $W_{d+1}''$ 
is given by the action evaluated on $W_{d+1}' \cup \overline W_{d+1}''$, where $ \overline W_{d+1}''$ is the orientation reversal of $W_{d+1}''$,\footnote{
	This is a general property of action which is local.
	Usually the locality is imposed by the requirement that
	an action $S$ evaluated on $W_{d+1}$ is given by an integral of a Lagrangian density as $S(W_{d+1}) = \int_{W_{d+1}} \cL$.
	However, this need not be the case in general.
	More general statement is that, an action $S$ satisfies
	$S(W_{d+1} \cup W_{d+1}') - S(W_{d+1}\cup W_{d+1}'') = S(W_{d+1}' \cup \overline W_{d+1}'') \mod 2\pi i$,
	where $W_{d+1} \cup W_{d+1}'$, $W_{d+1} \cup W_{d+1}''$, and $W_{d+1}' \cup \overline{W}_{d+1}''$ are all closed.
}
and the value of $2\pi i \int D \wedge d (C -A)$ is zero since $D=0$ on $W_{d+1}' \cup \overline{W}_{d+1}''$.
The value of $2\pi i \cdot Q(C)$ is also zero because $C=0$ on $W_{d+1}' \cup \overline{W}_{d+1}''$ and we have assumed that
$Q$ is determined by an element of $\Hom(\widetilde \Omega^{\text{spin}}_{d+1}(K(\bZ,4)) , \mathrm{U}(1))$. 
Notice that the reduced bordism group $\widetilde \Omega^{\text{spin}}_{d+1}(K(\bZ,4)) $ is used
rather than $\Omega^{\text{spin}}_{d+1}(K(\bZ,4)) $, and it is implicitly assumed that $Q(0)=0$.

\bigskip

As we have argued, there are no dynamical degrees of freedom inside the bulk. Therefore, all the degrees of freedom are localized near the boundary.
These degrees of freedom are described as follows.
For simplicity, let us first consider the case where the background field is set to zero, $A_e=0$.
We also assume that $Q(C)$ is either cubic in $C$ or topological 
so that it is irrelevant for the linearized equations of motion. Then the equations of motion in the Lorentzian signature metric (rather than the Euclidean
signature metric) is
\beq
(-1)^{d} \cdot d(*F_D) = 2\pi e'^2 \cdot F_C, \qquad d(*F_C) = 2\pi e^2 \cdot F_D,
\eeq
where $F_C := dC$ and $F_D := dD$ are the field strengths. 
Let $\tau \leq 0$ be the coordinate orthonormal to the boundary such that the boundary is located at $\tau =0$ and the bulk is in the region $\tau <0$.
The equations of motion have localized solutions of the form
\beq
F_C = d(e^{2\pi e e'\tau} ) \wedge F_B, \qquad F_D = \frac{e'}{e} \cdot d(e^{2\pi e e' \tau}) \wedge *_d F_B \label{eq:localsol}
\eeq
where 
$F_B$ is a 3-form which depends only on the coordinates of the boundary manifold $M_d$, and $*_d$ is the Hodge star
on the boundary.
The boundary condition \eqref{eq:boundarycondition} is satisfied since the differential form 
$d\tau$ becomes zero when it is pulled back to the boundary $\tau=0$.
These expressions for $F_C$ and $F_D$ are solutions of the equations of motion, if $F_B$ satisfies
\beq
dF_B =0, \qquad d(*_d F_B)=0.
\eeq
Therefore, $F_B$ is interpreted as the field strength of a 2-form field $B$ as $F_B = dB$,
where the 2-form fields are the boundary degrees of freedom. 
The above solution is exponentially localized near the boundary with the length scale $(2\pi e e')^{-1}$, so it is completely localized in the limit $ee' \to \infty$. 

When we turn on the background field $A_e$, one of the equations of motion is changed to $(-1)^{d} d(*F_D) = 2\pi e'^2 (F_C- F_{A_e})$, where $F_{A_e} = dA_e$.
Let us define a 3-form at the boundary by 
\beq
H:= \frac{ (-1)^{d}}{2\pi e'^2} \cdot *_{d+1} F_D|_{\tau =0}.
\eeq
Note that, although the pullback of $F_D$ to the boundary is zero by the boundary condition \eqref{eq:boundarycondition}, 
its Hodge dual $*F_D$ need not be zero at the boundary; indeed, if $A_e=0$, then $H = F_B = dB$ from the solution \eqref{eq:localsol}.
On the other hand, since the pullback of $F_C$ is zero at the boundary, we have
\beq
dH = -F_{A_e}, \label{eq:famousequation}
\eeq
meaning that $H$ can actually be written as $H = dB - A_e$.

\subsection{Anomaly cancellation}

Let us recapitulate the above results.
We introduced a theory which is defined on $(d+1)$-manifolds possibly with boundaries.
Inside the bulk, there are no dynamical degrees of freedom and the partition function is $2\pi i \cdot Q(A_e)$.
When boundaries exist, there is a localized degree of freedom which is namely a 2-form field $B$.
This means that the 2-form field on the $d$-dimensional boundary has the anomaly described by $Q(A_e)$.

Now we can discuss the anomaly cancellation.
Recall that the homotopy groups of the classifying space $BE_8$
are the same as those of the Eilenberg-MacLane space $K(\bZ,4)$ up to very high dimensions \eqref{homotopy-E6E7E8}, 
so that one can identify $K(\bZ,4) $ and $BE_8$ for the present purpose.
More concretely,
$E_8$-bundles on a manifold $X$ are classified by
the homotopy classes of classifying maps $f: X \to BE_8$,
and they correspond one-to-one with characteristic classes $f^* y \in H^4(X;\bZ) \simeq [X,K(\bZ,4)]$ 
associated with the generator $y \in H^4(BE_8; \bZ)$, if $\dim X < 15$.
This fact can be shown by obstruction-theoretic argument as represented in \cite{Witten:1985bt}.

Then, let us take the action $Q$ of $9d$ bulk to be the nontrivial element of 
\beq
\Hom(\widetilde \Omega^{\text{spin}}_{9}(K(\bZ,4)) , \U(1))  = \Hom(\widetilde \Omega^{\text{spin}}_{9}(BE_8) , \U(1)) = \bZ_2.
\eeq
If we take the background 3-form field $A_e$ on $X=\mathbb{HP}^2$ such that its 4-form flux $F_{A_e}$ is equal to the generator $x \in H^4(\mathbb{HP}^2;\bZ)$,
then $Q(A_e) = \frac 12 \mod \bZ$ on $\mathbb{HP}^2 \times S^1$. 
This is because the $E_8$ adjoint fermion had a nontrivial anomaly 
detected by the $E_8$-bundle $P_{E_8} \to \mathbb{HP}^2$ as seen in Sec.\,\ref{sec:explicit} (which corresponds to the nontrivial element of $\Hom(\widetilde \Omega^{\text{spin}}_{9}(BE_8) , \U(1)) = \bZ_2$),
and the characteristic class $f^* y \in H^4(\mathbb{HP}^2;\bZ)$
is equal to $x$ for the bundle $P_{E_8}$.
More generally, if the flux is $F_{A_e} = m x$ ($m \in \bZ$), then the anomaly is given by $Q(A_e) =\frac{m}{2} \mod \bZ$.

To cancel the anomaly of the adjoint fermion for generic $G$ detected by the bundle $P_G \to \mathbb{HP}^2$, we proceed as follows.
Take $A_e$ to be the Chern-Simons 3-form associated with the group $G$ such that
its restriction to $\SU(2)$ via $\SU(2) \to G$
gives a Chern-Simons 3-form for $\SU(2)$ with an odd level, where the map $\SU(2) \to G$ is the one used in the construction of the bundle $P_G$. 
This is always possible for simply-connected $G$, 
so suppose that $G$ is simply-connected for the moment. Then, we have $H^{i}(BG;\bZ) =0$ for $i<4$ and $H^4(BG;\bZ) = \bZ$, where the generator $c$ of the latter 
corresponds to an ``instanton number'' if we consider a classifying map $f : X \to BG$ and integrate the pullback $f^*c$ on a 4-manifold $X$. 
This ``instanton number'' of $G$ pulls back to that of $\SU(2)$ under
the map $\SU(2) \to G$, and thus
$c$ pulls back to the generator of $H^4(B\SU(2);\bZ)$.

The reason that we allow any odd Chern-Simons level $k_G$ is that our anomaly is $\bZ_2$ valued;
it must be odd for the anomaly of an adjoint fermion to be cancelled by 2-form fields. But note that, at the level of the present analysis, we can only determine it modulo 2.\footnote{
	It would be interesting to find a further restriction on $k_G$.
}
The level $k_G$ appears in the equation \eqref{eq:famousequation} where $F_{A_e}$ is now the 4-form constructed from the gauge field strength $F_G$,
\beq
dH = k_G \cdot \cN_G \tr (F_G \wedge F_G),
\label{eq:pure-gauge-CS}
\eeq
with
$\cN_G$ being an appropriate normalization factor such that $\cN_G \tr (F_G \wedge F_G)$
corresponds to the characteristic class $f^* c$ by the Chern-Weil construction.

Having chosen $A_e$ to be a Chern-Simons 3-form as above, we get an anomaly $Q(A_e)$ of the gauge group $G$ from the 2-form field.
By checking its value on the bundles $P_{G} \to \mathbb{HP}^2$ ($\times S^1$), we see that 
the anomaly of the fermion in the adjoint representation of $G$ is cancelled by $Q(A_e)$. 
More generally, we can explicitly check the anomaly cancellation for each generator manifold (equipped with $G$-bundle) of the bordism group.
Thus, we conclude that the fermion in the adjoint representation of $G = \SU(n), \Spin(2n), E_{6,7,8}, G_2$
can be cancelled by the 2-form.
(The situation is the same for a product group $G=G_1 \times G_2 \times \cdots$ of them.)

\bigskip

We remark that when $G$ is not simply-connected, it is not necessarily true that we can take such a generator $c\in H^4(BG;\bZ)$ which pulls back to the generator
of $H^4(B\SU(2);\bZ)$ \cite{Witten:2000nv,Cordova:2019uob}. 
The situation is similar for a more general gauge group 
\beq
G=\frac{G_1 \times G_2 \times \cdots \times H_1 \times H_2 \times \cdots}{Z}
\eeq
where $G_i$'s
are simple and simply-connected,
$H_j$'s
are groups whose adjoint fermions do not have the anomaly under discussion (such as $H_j=\U(1)$), and $Z$ is a center.
We want a Chern-Simons 3-form for $G$ such that if we restrict to $\SU(2)$ via $\SU(2) \to G_i \to G$,
we get an $\SU(2)$ Chern-Simons 3-form with an odd level. 
Such a Chern-Simons 3-form always exists if $Z$ is trivial, but more generally its existence depends on the global topology $\pi_1(G)$.
This point has been essentially discussed in \cite{Cvetic:2020kuw}, where it was found that this constraint (along with others)
gives very good agreement with the gauge groups explicitly realized in F-theory, at least for the case of rank $18$.

\bigskip

Finally let us incorporate a gravitino. It has a pure gravitational anomaly which is detected by $\mathbb{HP}^2$.
It can be cancelled as follows. In this paper we have been assuming that manifolds are spin, so the structure group
of the tangent bundle is $\Spin(d)$ in $d$-dimensions. This group has the Chern-Simons 3-form whose field strength
is half the first Pontrjagin class, $p_1/2$.
We add the Chern-Simons 3-form of this structure group to $A_e$ with an odd level. 
Since the first Pontrjagin class of $\mathbb{HP}^2$ is $p_1=2x$,
we have $p_1/2 = x$. Thus we see that the anomaly of the gravitino is cancelled in the same way as that of adjoint fermions
by replacing $f^\ast c$ with $p_1/2$. 
The equation for $H$ is now given by
\beq
dH = k_{\rm grav} \left[\cN_{\rm grav} \tr R \wedge R \right] + \sum_{i} k_{G_i}  \left[\cN_{G_i} \tr F_{G_i} \wedge F_{G_i}   \right] +\cdots \label{eq:Heq}
\eeq
where the notation is similar to \eqref{eq:pure-gauge-CS} and the ellipses denote possible terms which are not relevant for the present purposes.
 In particular, $k_{\rm grav}$ and $k_{G_i}$ are the Chern-Simons levels for the gravity and the gauge group $G_i$, respectively.
For the purpose of the anomaly cancellation by the 2-form field, we need to take them to be odd.

\section{Anomaly cancellation via topological degrees of freedom}\label{sec:TQFT}

In string theory, there are some situations in which
the mechanism discussed in Sec.\,\ref{sec:GSmech} is not sufficient to fully explain the anomaly cancellation.
Let us mention three examples. Actually, the first two of them are obtained by $S^1$ compactification of 9-dimensional theories,
so let us mention these 9-dimensional theories.

\begin{itemize}
\item M-theory on Klein bottle, or equivalently,
Type  IIA string theory on $S^1$ with a nontrivial holonomy of the $\bZ_2$ symmetry $(-1)^{F_L}$
which flips the sign of one of the two spinors of the 10-dimensional $\cN=(1,1)$ supersymmetry.
After the compactification, there is an $\cN=1$ supersymmetry in 9-dimensions and hence a single gravitino,
but $k_{\rm grav} = 0$.

\item M-theory on M\"obius strip, or equivalently,
$E_8 \times E_8$ heterotic string theory on $S^1$ with a nontrivial holonomy of the $\bZ_2$ symmetry
which exchanges two $E_8$'s.
After the compactification, we have a single $E_8$ gauge group in 9-dimensions, but $k_{E_8} = 2$
(which is the sum of the two Chern-Simons levels of the original $E_8$'s).

\item Type IIB string theory with three O7$^-$-planes, one O7$^+$-plane, and eight D7-branes on $T^2/\bZ_2$.
Putting $n$ D7-branes on top of the O7$^+$-plane, we get an $\Sp(n)$ gauge group.
As discussed in Sec.\,\ref{sec:explicit},
an adjoint fermion has an anomaly for $n\geq 2$ which is detected by the bundle $Q_G \to S^4 \times S^4$.
The subtlety of this anomaly was already discussed in \cite{Garcia-Etxebarria:2017crf}. 
\end{itemize}
In $8d$ $\cN=1$ supergravity, the only ranks of the total gauge group which are known to be realized in string theory
are $18$, $10$, and $2$~\cite{Montero:2020icj}.
The mechanism discussed in Sec.\,\ref{sec:GSmech} works in the case of rank $18$,
where all known examples have odd $k_{\rm grav}$ and $k_G$,
while the first example above is the case of rank $2$, and
the second and third examples are the cases of rank $10$.

Here we focus our attention on the first example mentioned above.
We argue that there is a topological degree of freedom, namely a 3-form $\bZ_2$ gauge field, which cancels the anomaly of a gravitino. 

\subsection{Topological degrees of freedom}

To see the topological degree of freedom and its effect on the topology of spacetime, we first recall some facts about M-theory~\cite{Witten:1996md,Freed:2019sco}. 
M-theory contains a 3-form field $C$, and its 4-form flux $G$
is known to satisfy the shifted quantization condition~\cite{Witten:1996md}.
Let $[2G]_2 \in H^4(X_{11}; \bZ_2)$ be a mod-2 reduction of an (orientation-bundle twisted) integral cohomology class $2G \in H^4(X_{11};\widetilde \bZ)$.
Then we have
\begin{equation}
	[2G]_2 = w_4,
	\label{eq:mc}
\end{equation}
where $w_4 \in H^4(X_{11}; \bZ_2)$ is the fourth Stiefel-Whitney class.
Also, M-theory has parity (or orientation-reversal) symmetry,
under which $C$ is odd and the sign is flipped, $C \to -C$ (and correspondingly $G \to -G$).

Now, consider a manifold $X_{11} = M_9 \times \KB$, 
where $M_9$ is a 9-dimensional spin manifold and $\KB$ is a Klein bottle.
The Kaluza-Klein reduction of $C$ in the $\KB$ compactification contains components $C_{\mu\nu\rho}$ which are independent of the coordinates of $\KB$
and whose three indices are in the direction of $M_9$. These components become a 3-form field on $M_9$ which we also denote as $C$ by abuse of notation.
However, this 3-form $C$ on $M_9$ is severely constrained.
By going around a loop in $\KB$ along which the orientation is reversed, 
the sign of $C$ is flipped since $C$ is parity-odd.
On the other hand, $C$ is independent of the coordinates of $\KB$.
These two facts conspire to conclude $C = -C$ up to gauge transformation, which then imply $G = - G$ or equivalently $2G=0$.

From \eqref{eq:mc}, this means that the condition $w_4=0$ must be imposed on 9-dimensional manifolds $M_9$ in the low energy theory 
after the compactification on $\KB$. In general, we believe (although do not show generally) that such a condition cannot be ``put by hand''.
For example, in the case of the previous section, an analogous topological condition is that the right hand side of \eqref{eq:Heq} is cohomologically trivial;
this is not imposed by hand, but is realized by the 2-form field $B$. In a similar way, locality presumably requires that the
condition $w_4=0$ is imposed by a 3-form $\bZ_2$ gauge field, which is described as follows. Let $\mathsf{w}_4 \in Z^4(M_9;\bZ_2)$ 
be an explicit cocycle representing $w_4$, i.e. $w_4 = [\mathsf{w}_4]$. 
Then the fact that $w_4$ is trivial means that there is another cochain $\mathsf{v}_3 \in C^3(M_9;\bZ_2)$
such that
\beq
\delta \mathsf{v}_3 = \mathsf{w}_4, \label{eq:trivialization}
\eeq
where $\delta$ is the coboundary operator.
This equation is analogous to \eqref{eq:Heq}. It does not completely specify $\mathsf{v}_3$. Indeed,
let $\mathsf{v}'_3$ be another cochain satisfying the same equation. Then we have $\delta (\mathsf{v}_3 - \mathsf{v}'_3)=0$
and hence there is an ambiguity of $\mathsf{v}_3 - \mathsf{v}'_3 \in Z^3(M_9;\bZ_2)$.  This gives some topological degrees of freedom.
It is likely that we should impose a gauge equivalence condition $\mathsf{v}_3 \sim \mathsf{v}_3 + \delta \mathsf{u}_2$
for cochains $\mathsf{u}_2 \in C^2(M_9; \bZ_2)$. If so, the degrees of freedom contained in $\mathsf{v}_3$ is described by $H^3(M_9; \bZ_2)$.
This kind of structure is called the (degree-4) Wu structure \cite{Monnier:2016jlo, Monnier:2018nfs}.

We remark that if we replace $w_4$ by $w_2$ and consider oriented manifolds instead of spin manifolds in the above discussion, 
the corresponding (degree-2) Wu structure would be a spin structure.
In that case, $w_2=0$ implies the existence of a spin structure, and a choice of $\mathsf{v}_1$ such that $\delta \mathsf{v}_1 = \mathsf{w}_2$
corresponds to a choice of an explicit spin structure.\footnote{
The fact that $\mathsf{v}_1$ corresponds to a spin structure can be seen as follows.
Consider the $\SO(d)$ bundle associated with the tangent bundle of the manifold, and let $g_{ij}$ be transition functions
between two patches $U_i$ and $U_j$ which take values in $\SO(d)$. Letting $\hat{g}_{ij}$ be a lift of $g_{ij}$ to $\Spin(d)$,
the cocycle $(\mathsf{w}_2)_{ijk}$ may be defined as $\hat{g}_{ij}\hat{g}_{jk}\hat{g}_{ki} = (-1)^{(\mathsf{w}_2)_{ijk}}$.
If we define $\tilde{g}_{ij} = (-1)^{(\mathsf{v}_1)_{ij}} \hat{g}_{ij}$, we get
$\tilde{g}_{ij}\tilde{g}_{jk}\tilde{g}_{ki} = 1$ and it gives a $\Spin(d)$ bundle. Thus, a choice of $\mathsf{v}_1$ gives a spin structure.
}
Mere the existence of a spin structure is not enough for locality; we need explicit spin structures on manifolds.\footnote{
One can find two manifolds $N_d$ and $N'_d$
with a common boundary $\partial N_d = \partial N'_d$
such that spin structure exists on $N_d$ and $N'_d$, but not on $N_d \cup \overline{N'_d}$. 
Thus, ``the existence of a spin structure'' (rather than explicit spin structure) is not a local concept. For example, take $N_4$ to be a half K3 surface with the boundary $T^3$
(which is obtained by e.g. an elliptic fibration over a hemisphere),
and $N'_4$ to be $D^2 \times T^2$. These manifolds $N_4$ and $N'_4$ can be glued without any problem if we do not care about spin structures, 
but they cannot be glued keeping spin structures consistent. A simpler example in the case of pin$^+$ structure rather than spin structure is to take $N_2$ to be a crosscap with the boundary $S^1$, and $N'_2$ to be $D^2$.
By gluing them, we get a real projective space $\mathbb{RP}^2$ which does not possess a pin$^+$ structure.
}

In the present situation of M-theory compactified on $\KB$, 
there is a perfect candidate for such a 3-form $\bZ_2$ gauge field. We have discussed that the consistency requires that $C = -C$ or $2C=0$ on $M_9$
up to gauge transformations.
This does not imply $C=0$; rather, it implies that $C$ is torsion. 
Thus $C$ itself is a 3-form $\bZ_2$ gauge field on $M_9$.
More explicitly, when it is integrated over 3-cycles, it takes values $0$ or $\frac12$ mod $\bZ$. 
Thus we can identify 
\beq
C \sim \frac{1}{2}\mathsf{v}_3 \mod \bZ,
\eeq
where we only consider modulo $\bZ$ corresponding to the gauge equivalence. 
Although we do not try to make all mathematical details precise,\footnote{For a more precise definition of $C$,
one must consider the M2-brane partition function and its anomalies~\cite{Witten:2016cio}. 
The requirement is as follows. Let $N_3$ be the worldvolume of an M2-brane, and $Z(N_3)$ be the (anomalous) partition function
of the degrees of freedom on the M2-brane. Then the product $Z(N_3) \exp(2\pi i \int_{N_3} C)$ must be well-defined.
} 
this identification suggests the desired result \eqref{eq:trivialization} 
since we may think that $G \sim \delta C$ and hence $2G \sim \delta \mathsf{v}_3$, while we also had $2G \sim \mathsf{w}_4 $ in \eqref{eq:mc}.

\subsection{Anomaly cancellation}
We have discussed the existence of topological degrees of freedom $\mathsf{v}_3$ which is an explicit trivialization
of the cocycle $\mathsf{w}_4$ representing $w_4$. 
Now we would like to discuss how it is relevant for the anomaly cancellation. 
For this purpose, we use the results of \cite{Witten:1996md,Freed:2019sco}.
It was found there that the gravitino in 11-dimensional supergravity has an anomaly, but this anomaly can be cancelled by 
a cubic Chern-Simons term of the 3-form $C$, which is roughly $\frac{1}{6} C \wedge G \wedge G + I_8(R) \wedge C$ where $I_8(R)$
is an 8-form constructed from the Riemann curvature $R$.
The anomaly of the 11-dimensional gravitino is represented by a 12-dimensional invertible field theory.
Although this is nonzero, it is equal (with the opposite sign) to the integral of the 12-form $\frac{1}{6} G \wedge G \wedge G + I_8(R) \wedge G$
as far as the 4-form $G$ satisfies the condition \eqref{eq:mc}. Thus the sum of the anomaly of the gravitino and the integral of this 12-form is zero.

Now let us restrict our attention to the case $Y_{12} = W_{10} \times \KB$, where $W_{10}$ is a spin manifold with $w_4=0$.
The Klein bottle also has $w_4=0$, and we can take $G=0$ consistently with the condition \eqref{eq:mc}.
Then the contribution from the 12-form is zero, and hence the contribution from the gravitino must be also zero 
according to the results of \cite{Witten:1996md,Freed:2019sco}. 
Our theory is obtained by the dimensional reduction on $\KB$, so we conclude that the evaluation of the gravitino anomaly on manifolds $W_{10}$
with $w_4=0$ is zero. 

What we have found above is the fact that the anomaly of the gravitino is zero 
if the theory is formulated in the bordism category of manifolds with Wu structure. 
After the explicit path integral over the topological degrees of freedom $\mathsf{v}_3$,
the Wu structure is ``integrated over'' and we expect to get a topological quantum field theory (TQFT) 
which is defined in the bordism category of manifolds with spin structure.
This is analogous to the situation that the sum over spin structures give a ``bosonic'' (non-spin) theory which does not depend on spin structure.
Now the question is whether the (non-Wu) spin-TQFT reproduces the anomaly of the gravitino. 
The construction of a class of TQFTs relevant to the current situation been given in \cite[Sec.\,2.4]{Kobayashi:2019lep}.
The anomaly of the TQFT coupled to a background $\bZ_2$ 4-form field is 
classified by $\widetilde\Omega_9^{\text{spin}}(K(\bZ_2, 4))$, and this anomaly trivializes when the background is turned off.
The construction of \cite{Kobayashi:2019lep} works in this kind of situation.
We have found in Sec.~\ref{sec:Z2} and \ref{sec:str} that
the group $\widetilde\Omega_9^{\text{spin}}(K(\bZ_2, 4))$ contains an element represented by $\mathbb{HP}^2 \times S^1$ with a nontrivial background of $H^4(\mathbb{HP}^2;\bZ_2)$ turned on.
By taking the background $\bZ_2$ 4-form to be $w_4$, we get the desired anomaly which cancels against the gravitino anomaly.

We leave the rank 10 cases mentioned at the beginning of this section 
(e.g. the case of $E_8$ with level 2 and the case of $\Sp(n)$) for future work.
Some of the rank 10 theories are constructed in heterotic string theories~\cite{Chaudhuri:1995fk}, and the results
of \cite{Tachikawa:2021mby} suggests that fermion anomalies may be zero as long as the 2-form field is regarded as a background
field imposing the (twisted) string structure \eqref{eq:Heq}. The explicit path integral over the 2-form field is subtle,
but the appropriate action for the 2-form field may be obtained by the construction along the lines of \cite{Kobayashi:2019lep,Yonekura:2020upo}.
It is important to understand what (topological) degrees of freedom exist in the theory.
The same question also arises in Type~IIB string theory~\cite{Debray:2021vob}.

\bigskip

\section*{Acknowledgments}
The authors would like to thank Kantaro Ohmori for collaboration in the early stage of this work,
and Yuji Tachikawa for helpful discussions and comments.
YL is partially supported by the Programs for Leading Graduate Schools, MEXT, Japan,
via the Leading Graduate Course for Frontiers of Mathematical Sciences and Physics
and also by JSPS Research Fellowship for Young Scientists.
The work of KY is supported in part by JST FOREST Program (Grant Number JPMJFR2030, Japan), 
MEXT-JSPS Grant-in-Aid for Transformative Research Areas (A) ”Extreme Universe” (No. 21H05188),
and JSPS KAKENHI (17K14265).

\newpage

\bibliographystyle{ytamsalpha}
\baselineskip=.95\baselineskip
\bibliography{ref}

\end{document}